\documentclass[times,twocolumn,final]{elsarticle}

\usepackage{medima}
\usepackage{framed,multirow}

\usepackage{amssymb}
\usepackage{latexsym}
\usepackage{amsmath}

\usepackage{url}
\usepackage{xcolor}

\usepackage{hyperref}

\definecolor{newcolor}{rgb}{.8,.349,.1}

\journal{Medical Image Analysis}

\begin{document}

\verso{Elforaici \textit{et~al.}}

\begin{frontmatter}

\title{Semi-supervised ViT knowledge distillation network with style transfer normalization for colorectal liver metastases survival prediction}%


\author[1,2]{Mohamed El Amine \snm{Elforaici} \corref{cor1}}
\ead{maelforaici@polymtl.ca}
\author[2]{Emmanuel \snm{Montagnon}}
\author[1,2]{Francisco \snm{Perdigón Romero}}
\author[1,2]{William Trung \snm{Le}}
\author[2]{Feryel \snm{Azzi}}
\author[2,3]{Dominique \snm{Trudel}}
\author[3]{Bich \snm{Nguyen}}
\author[2,4]{Simon \snm{Turcotte}}
\author[2,5]{An \snm{Tang}}
\author[1,2,3]{Samuel \snm{Kadoury}}

\address[1]{MedICAL Laboratory, Polytechnique Montr\'eal, Montreal, Canada}
\address[2]{Centre de recherche du CHUM (CRCHUM), Montreal, Canada}
\address[3]{Universit\'e de Montr\'eal, Montreal, Canada}
\address[4]{Department of surgery, Universit\'e de Montr\'eal, Montreal, Canada}
\address[5]{Department of Radiology, Radiation Oncology and Nuclear Medicine, Universit\'e de Montr\'eal, Montreal, Canada}

\received{30 May 2023}

\begin{abstract}
Colorectal liver metastases (CLM) affect almost half of all colon cancer patients and the response to systemic chemotherapy plays a crucial role in patient survival. While oncologists typically use tumor grading scores, such as tumor regression grade (TRG), to establish an accurate prognosis on patient outcomes, including overall survival (OS) and time-to-recurrence (TTR), these traditional methods have several limitations. They are subjective, time-consuming, and require extensive expertise, which limits their scalability and reliability. Additionally, existing approaches for prognosis prediction using machine learning mostly rely on radiological imaging data, but recently histological images have been shown to be relevant for survival predictions by allowing to fully capture the complex microenvironmental and cellular characteristics of the tumor. To address these limitations, we propose an end-to-end approach for automated prognosis prediction using histology slides stained with Hematoxylin and Eosin (H\&E) and Hematoxylin Phloxine Saffron (HPS). We first employ a Generative Adversarial Network (GAN) for slide normalization to reduce staining variations and improve the overall quality of the images that are used as input to our prediction pipeline. We propose a semi-supervised model to perform tissue classification from sparse annotations, producing segmentation and feature maps. Specifically, we use an attention-based approach that weighs the importance of different slide regions in producing the final classification results. Finally, we exploit the extracted features for the metastatic nodules and surrounding tissue to train a prognosis model. In parallel, we train a vision Transformer model in a knowledge distillation framework to replicate and enhance the performance of the prognosis prediction. We evaluate our approach on a clinical dataset of 258 CLM patients, achieving superior performance compared to three other models with a c-index of 0.804 (0.014) for OS and 0.733 (0.014) for TTR. The proposed approach achieves an accuracy of 86.9\% to 90.3\% in predicting TRG dichotomization. For the 3-class TRG classification task, the proposed approach yields an accuracy of 78.5\% to 82.1\%, outperforming the comparative method. Our proposed pipeline can provide automated prognosis for pathologists and oncologists, and can greatly promote precision medicine progress in managing CLM patients.
\end{abstract}


\end{frontmatter}


\section{Introduction}
Colorectal liver metastasis (CLM) is a common and deadly condition in which cancer cells from the colon or rectum spread to the liver \cite{xi2021global}. Treatment selection for CLM requires a thorough understanding and characterization of cancer in the patient to determine prognosis and identify appropriate treatment options. While traditional staging methods such as the clinical risk score (CRS) \cite{fong1999clinical} or the tumor regression grade (TRG) \cite{mandard1994pathologic} have been developed to classify patients into low- and high-risk groups that would respond to treatments differently, there is significant variation in patient outcomes even within a single TRG or CRS. Thus, more accurate patient risk classification is needed to improve patient management and disease outcomes.

In recent years, there has been a surge of interest in using machine learning (ML) techniques to provide unique prognostic information from histopathology imaging, that complements the most recent clinical recommendations. However, obtaining insights on machine-learned prognostic features remains challenging despite some existing efforts \cite{van2021deep}. The identification of potentially innovative traits and the establishment of interpretable information required for AI-supported clinical decision-making  could be made possible if the learned features can be reliably detected and shown to have independent prognostic value. Additionally, the annotation of data in a clinical context is essential for training ML models but remains a time-consuming and expensive task \cite{boehm2022harnessing}.

Previous deep learning-based efforts to predict clinical outcomes using histopathology slides can be divided into two main categories. First, using specialized tools like CellProfiler \cite{mcquin2018cellprofiler}, these approaches concentrate on extracting predefined morphological features from the slides. Then, they use statistical or machine learning methods to determine which of the predefined features are connected with survival or recurrence. In the second category, techniques avoid the need to extract predefined features by using weakly-supervised DL approaches to predict survival directly from whole slide images (WSI) \cite{wulczyn2020deep}. However, to be suitable in a clinical context, the histology slides used in these applications must be appropriately standardized. Since multiple staining techniques are used by pathologists to visualize certain tissue characteristics, where solutions and procedures within one staining technique can differ from one center to another, stain normalization of slides is necessary, particularly when training deep neural networks for image classification or segmentation, to make the models transferable to other datasets \cite{ciompi2017importance}.

However, despite the potential benefits of machine learning in the field of pathology, its use remains limited due to a lack of interpretability and transparency. Pathologists require confidence in the accuracy of machine learning models, particularly in applications such as cancer prognosis, where erroneous prognosis can have severe consequences \cite{border2022growing}. Thus, there is a need for ML models that can not only make accurate predictions but also provide explanations for their decisions, highlighting the regions of interest that contribute most significantly to the prediction \cite{banegas2021towards}. 

In this paper, we propose a semi-supervised learning (SSL) approach for colorectal liver metastases based on histopathological images with a visual Transformer for distilling knowledge to a student-teacher framework. The paper's contributions are the following:

\begin{enumerate}
\item We introduce a generative style transfer approach into WSI normalization to exploit slides stained with two different methods: hematoxylin and eosin (H\&E) and hematoxylin phloxine saffron (HPS). 

\item We propose a semi-supervised deep learning (DL) algorithm based on a mean-teacher approach for classifying digital surgical resection slides into five classes: normal tissue, fibrosis, cancer, necrosis, and background. We then predict disease-specific survival and time to recurrence using normalized H\&E and HPS slides combined with clinical data.

\item We integrate a vision Transformer to distill knowledge from the trained DL model into a smaller and more computationally efficient model that can be used for inference on new data. By doing so, we are able to reduce the computational cost of our model while maintaining high accuracy and providing interpretability through the use of the vision Transformer.

\item We evaluate the performance of our proposed model on an in-house dataset of 258 colorectal cancer patients, compared to three other state-of-the-art methods, providing extensive experimental results to assess the performance in its prediction capacity.
\end{enumerate}

\section{Related works}

\subsection{Image normalization}
Various advanced techniques exist in the literature for standardizing colors in histopathological images, including stain separation, template color-matching, and style transfer using generative models. Using pixel level statistical color descriptors, some previous works proposed supervised methods to measure the stain concentration matrix (SCD) \cite{zheng2020stain}. They performed normalization of the source image to the target image's color space via a nonlinear color mapping process. Compared to other state-of-the-art procedures, these methods demand higher computational complexity. In LAB color space, Reinhard et al. presented a color mapping technique where each color channel of the source image is matched with the color channel of the user selected template image \cite{reinhard2001color}. The standardized images were then transformed to a different color space to perform normalization, and then back to RGB. This color matching method makes the assumption that each dyeing agent's fraction of tissue components is consistent across the samples. As a result, this approach results in incorrect color matching because the white background is mapped as a colorful region \cite{vijh2021new}.
Another approach developed by Macenko et al. locates the optical density space's singular value decomposition (SVD) values and projects the data onto the plane that corresponds to the two biggest singular values. This method exhibits low computational complexity and can be used with other histology stains \cite{macenko2009method}.
The stain color adaptive normalization (SCAN) algorithm was recently developed as an unsupervised normalizing method. Segmentation and clustering techniques serve as the foundation for the SCAN algorithm \cite{salvi2020stain}.
On the other end of the spectrum, distinct from standard color deconvolution methods, there exist various stain transfer methods based on Generative Adversarial Networks (GAN). Although the simple GAN network performs well on natural images, it is unable to preserve the structural contents in images of histology \cite{liang2020stain, runz2021normalization}. The GAN-based techniques employ a color transfer procedure involving a collection of images. As a result, they effectively learn dataset specific features but disregard color patterns unique to individual images in the whole slide image (WSI).

\subsection{Survival prediction}
In recent years, several approaches for predicting survival using pathological slides have been proposed. They can be divided into two categories: ROI-based approaches and WSI-based methods.

\subsubsection{ROI-based analysis}
While the high resolution of HPS images could substantially aid in survival analysis, this property unfortunately makes the majority of existing models and algorithms computationally intractable on pathology sample inputs. The majority of methods previously focused on regions of interest (ROI) patches that pathologists chose from WSIs due to a lack of computational resources \cite{cui2021artificial}.

In an effort to produce reliable prognostic information, Yu et al. extracted 9879 quantitative image features from annotated ROIs \cite{yu2016predicting}. The results show that these features can predict the prognosis of lung cancer patients. Yao et al. went beyond conventional cell identification by first classifying various cell subtypes using a deep subtype cell detection method, and then retrieved features from the cellular subtype data \cite{yao2020whole}. In order to describe cell type distributions using ROIs for prediction, Cheng et al. employed a deep autoencoder to group cell patches into several types \cite{wang2019weakly}. Based on nucleus detection and segmentation, these algorithms extracted hand-crafted features that were thought to convey prior understanding of border, region, or shape. However, hand-crafted features are limited in representation power and scalability.

Deep learning-based survival models on the other hand have recently been presented for discovering more powerful representations from different kinds of data \cite{wulczyn2020deep, wulczyn2021interpretable, vale2021long, deepa2022systematic}. To describe the nonlinear risk function, Katzman et al. first developed a deep fully connected network (DeepSurv) \cite{katzman2018deepsurv}. They showed that DeepSurv performed better than the conventional linear Cox proportional hazard model.
DeepConvSurv, on the other hand, was proposed  to leverage the pathologists' preselected ROI patches from WSIs for convolution operations \cite{zhu2016deep}. A limited number of image tiles might not accurately and totally depict the tumor form of the patient. These techniques also employ average pooling to produce patient-wise predictions using patch-based data. Such a combination requires more consideration because it cannot efficiently aggregate predictions at the patch level. Therefore, it would be highly beneficial to discover deep knowledge from large complete slide images.

\subsubsection{Whole Slide Image survival analysis}
A number of methods for WSI analysis have been presented for a variety of applications, such as classification, detection, or segmentation, with detailed and densely annotated WSIs \cite{dimitriou2019deep, khened2021generalized, li2022comprehensive}. DL has shown potential when used for supervised learning in computational pathology.
It is a painstaking task to manually annotate large amounts of data, which makes it unfeasible in clinical practice. Additionally, the success of these applications depends on combining labor-intensive annotations and comprehensive patch contents, which might not be appropriate for survival prediction \cite{van2021deep}.
Weakly supervised methods could be one approach to effectively overcome the inadequacies of present models. Many weakly supervised medical image algorithms have recently been proposed by different research groups. The most distinct regions that correlate to different tumor types can be found using WSI classification and segmentation models \cite{campanella2019clinical, roth2021going, ren2023weakly}.

Furthermore, current WSI classification tasks are weakly supervised and based on slide-level data, whereas patient-level data is used to predict survival (one patient might have multiple whole slide images). The goal of these works is not to produce patient-level decisions based on data at the slide level but to provide a better understanding of the underlying patterns and features of WSIs that may contribute to the prediction of patient clinical outcomes such as survival. Zhu et al. suggested a patch-based two-step framework to predict patients' survival outcomes from WSI without the use of annotations. In the first stage, patches are extracted from the WSIs and clustered into various patterns known as "phenotypes" based on how they appear visually. Subsequently, WSISA \cite{zhu2017wsisa} used DeepConvSurv to choose crucial patch clusters, which were then aggregated for the outcome prediction.

More recently, DL has witnessed a surge in self-supervised learning as a paradigm for learning feature representations without using any labels. In order to learn meaningful representations of high-dimensional data, self-supervised learning makes use of auxiliary tasks such as recognizing that the representation of an image should not change significantly. There are two limitations of this current line of work, despite the fact that self-supervised learning has been suggested as an alternative for ResNet-50 encoders pre-trained on ImageNet \cite{deng2009imagenet} in pathology. First, there are few comprehensive benchmarks available for testing self-supervised models on various patch-level and weakly supervised tasks. The second limitation is related to the absence of introspection and posthoc evaluation of the acquired self-supervised representations, which obstructs the identification of the learned morphological features.\cite{chen2022self}.

\subsection{Knowledge distillation}
Knowledge distillation (KD) is a technique seeking to compress a large, complex model into a smaller, simpler model while enhancing its performance. A study by Romero et al. introduced knowledge distillation as a method to transfer knowledge from a larger model to a smaller one \cite{romero2014fitnets}. They demonstrated that their method was effective in improving the accuracy of smaller models.
Another study by Zheng et al. proposed a method for boosting contrastive learning with relation knowledge distillation \cite{zheng2022boosting}. Their method combined knowledge distillation with contrastive learning to improve the performance of a deep neural network for image classification. The results showed that their method outperformed existing methods for image classification.

\subsection{Vision Transformers}
Finally, deep neural network models known as vision transformers (ViT) have become highly popular in computer vision tasks because of their excellent performance in image classification, segmentation, and object detection tasks. The model architecture relies on self-attention mechanisms to compute feature representations of the input image and learn long-range dependencies between pixels \cite{dosovitskiy2020image}.

One promising application of ViT is in pathology, where the ability to accurately detect and classify abnormalities on histology slides can have a significant impact on patient outcomes. For example, a recent study used a ViT model to classify multiple patterns of images extracted from the WSI of a prostate biopsy, outperforming traditional CNNs \cite{ikromjanov2022whole}.

All the studies mentioned above demonstrate the importance of image normalization, prognosis prediction, knowledge distillation, and Vision transformers in histopathological image analysis. Our work builds upon these techniques to explore and improve the accuracy and efficiency of diagnosis and prognosis prediction using WSI.

\section{Materials and methods}

\begin{figure*}[t]
    \centering
    \includegraphics[width=19cm]{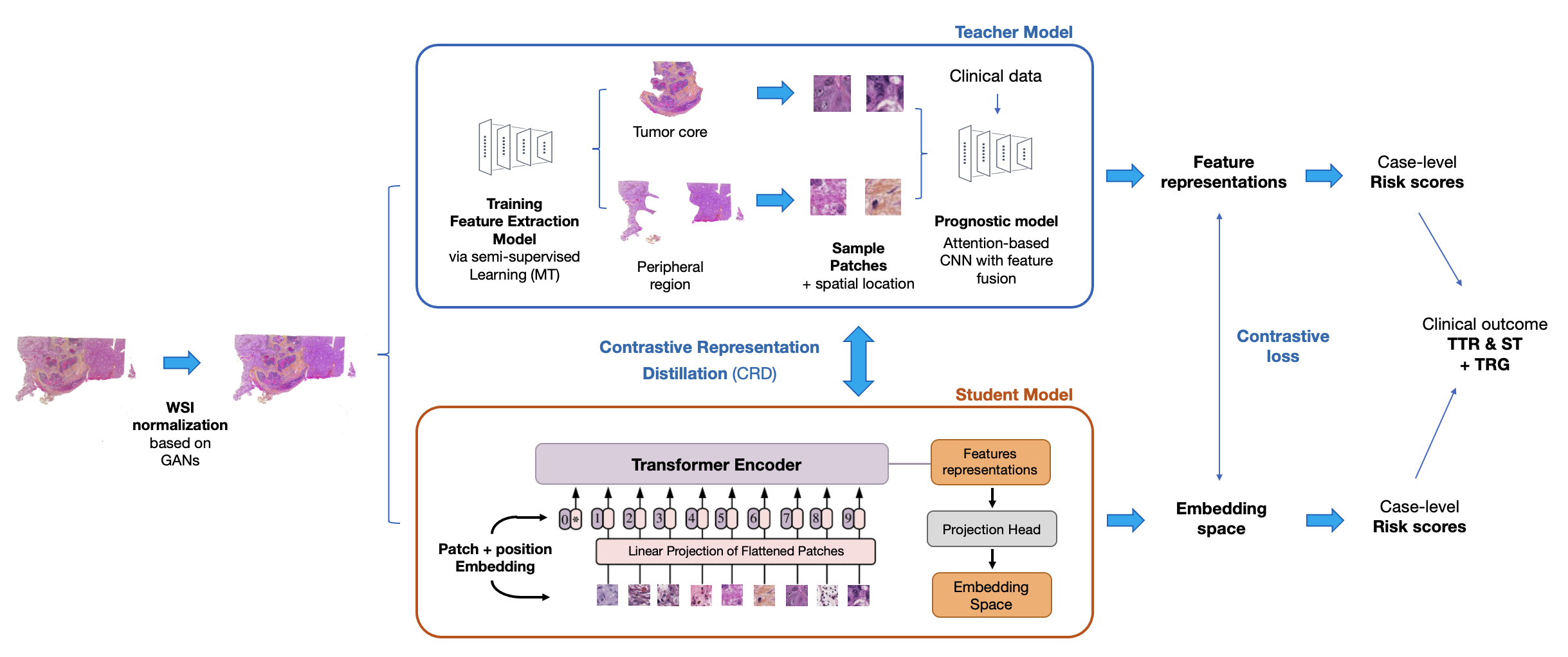}
    \caption[fig1]
    {\label{fig1} Overview of the semi-supervised ViT knowledge distillation network. First, the WSIs are normalized using the GAN model. Then, a Mean Teacher (MT) approach is trained to extract key features related to prognostic from the selected ROIs: tumor core and peripheral region. This part of the model is capable of doing the tissue classification task and will generate the classification maps. Using the extracted features from sample patches from the tumor core and the peripheral region, we train the Prognosis model through an attention mechanism with clinical data. In parallel, Contrastive Representation Distillation (CRD) uses a contrastive learning objective to encourage the student model (ViT) to learn similar representations to the teacher model, without relying on a direct feature matching objective.}
\end{figure*}

In this section, we present our overall framework illustrated in Figure \ref{fig1}. In Sec. 3.2, we present the normalization method based on GAN to normalize H\&E and HPS slides. Sec. 3.3 presents the Semi-supervised ViT knowledge distillation network that achieves the prognosis prediction, including overall survival (OS), time-to-recurrence (TTR), TRG and survival time (ST). Sec. 3.4 presents the experimental setup.

\subsection{Datasets and data preparation}
\label{ssec:dataset}
In this work, we used an unidentifiable and pseudo-anonymized dataset from a hepatobiliary biobank from the Montreal University Hospital Center (CHUM). It contains 1620 histological slides from 258 patients with CLM treated with chemotherapy and who subsequently underwent surgical resection. The cohort's clinical details can be summarized as follows: Out of a total of 258 individuals, men accounted for 60.5\% (156/258) while women represented 39.5\% (102/258). The average age of the participants was 64.8 years with a standard deviation of 10.2 years. On average, individuals had 3.7 liver metastases, with a standard deviation of 2.8. The maximum TRG scores were distributed as follows: 35 with TRG 1, 42 with TRG 2, 68 with TRG 3, 988 with TRG 4 and 15 with TRG 5.

H\&E and HPS stained histology slides were scanned with a NanoZoomer-XR scanner from Hamamatsu \cite{Hamamatsu} and stored in the Hamamatsu NanoZoomer Digital Pathology Image (NDPI) file format. In order to visualize the slides and annotate them, we used the NDP.view2 software \cite{NDPview2}. The five-classes annotation task was made on 147 slides by a histopathologist with an annotated area that did not exceed 1\% of the total area of the slide, and the five classes are: normal tissue, fibrosis, cancer, necrosis, and background. Handling data formats and conversions were performed using the OpenSlide library \cite{goode2013openslide}. 
In addition to these high-quality images, images were matched with a clinical dataset containing TRG values associated with each slide was used. These values were attributed by expert pathologists in the clinical context. When the TRG was not available, a pathologist reviewed the case to assign the TRG using a 5-point grading system proposed by Rubbia-Brandt \cite{rubbia2007importance}, where TRG 1 corresponds to a complete response with an absence of residual cancer, TRG 2 to the presence of residual cancer cells scattered through the fibrosis, TRG 3 to an increase in the number of residual cancer cells, with fibrosis predominant, TRG 4 to a residual cancer outgrowing fibrosis, and TRG 5 to the absence of regressive changes \cite{vecchio2005relationship}. Of the 1620 slides, 135 were TRG 1, 156 were TRG 2, 551 were TRG 3, 712 were TRG4 and 66 were TRG 5. In addition, we had for each patient the disease-specific survival time (ST) defined as the time interval between the date of metastases resection surgery and the cancer-related death, and TTR defined as the time interval between the date of metastases resection surgery and the first recurrence diagnosis.
In cases where patients did not experience recurrence or cancer-related death, their time of observation was considered to be the date of their last follow-up, which was censored for this study.

Following a 10$\times$ magnification factor, slides were converted from the NDPI format to the RGB format using the OpenSlide library. This allowed the transfer from a space containing the histology slide formatted in a pyramidal stack of different resolutions to a 2D image with a single resolution. Images were then normalized using the proposed GAN model.

During training, 32$\times$32 pixels patches were extracted from the slides and annotated based on the manually drawn box position in the slide. Patches were divided into two groups: (1) annotated patches retrieved from a bounding box of a certain color depending on the tissue class and (2) unannotated patches. In order to compensate for the class imbalance, especially between normal tissue and cancer, we applied a data augmentation strategy that consists of a 90\textdegree~ rotation and horizontal flipping.

\subsection{Normalization framework}
\begin{figure*}[t]
    \centering
    \includegraphics[width=\textwidth]{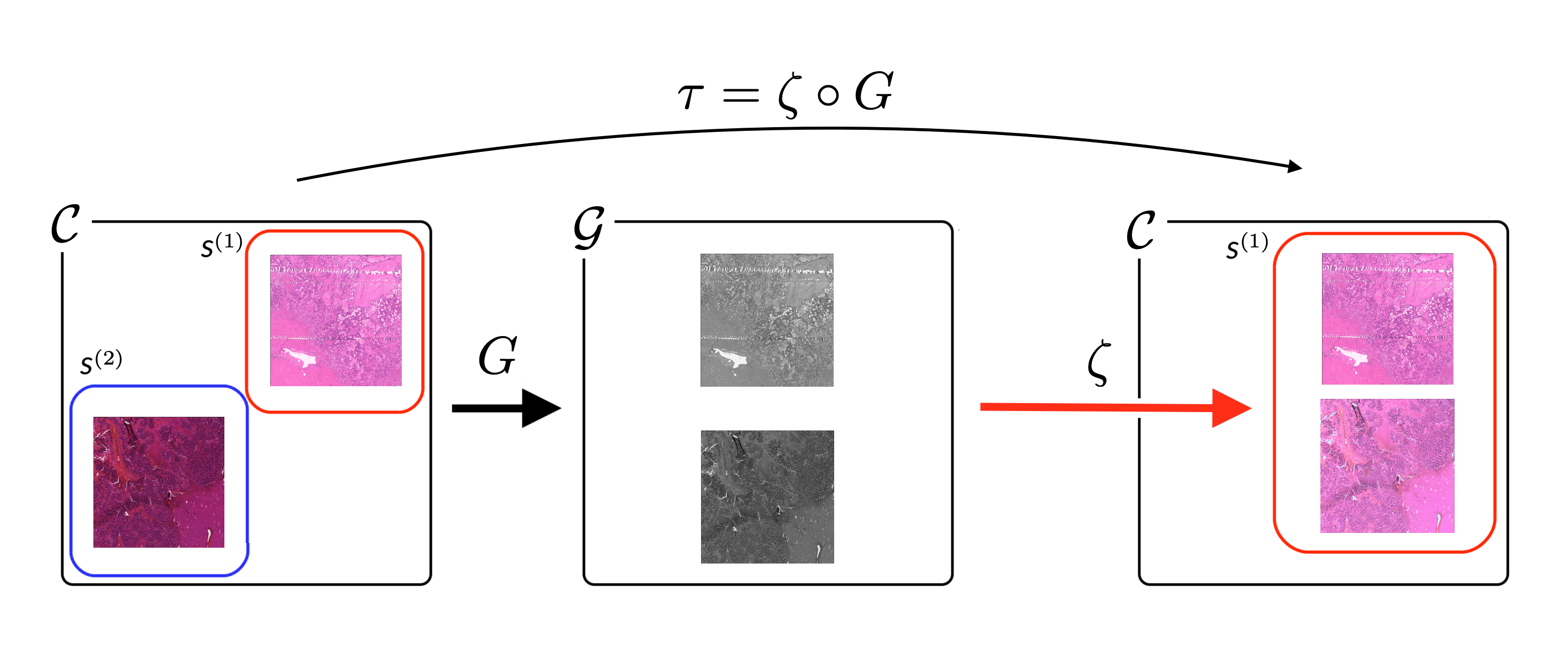}
    \caption[fig2]
    {\label{fig2} Overview of the stain-style transfer model. The model $\tau$ is composed of two transformations: Gray-normalization $\it{G}$ and style-generator $\zeta$. $\it{G}$ standardizes each stain-style, H\&E and HPS, and $\zeta$ colorizes gray images following the stain-style chosen as reference, in this case H\&E.}
\end{figure*}

The proposed normalization workflow of WSI is shown in Figure \ref{fig2}. The dataset contains slides stained with two different processes: H\&E which is considered the standard stain, and HPS. In order to use all the slides in the dataset, a generative model was proposed to normalize the color staining of the HPS slides, based on the stain-style transfer approach.

An iterative process allows the generative model, composed of a generator and a discriminator, to learn and improve over multiple cycles of image normalization. The task of the generator can be intuitively described as follows. In each step, the generator produces first a gray image with $G$ from the original image, then colorizes this image with style generator $\zeta$. The discriminator is presented with a few "real" data examples of selected H\&E slides, together with the examples produced by the generator, and its task is to classify them as “real” or “fake” depending on the normalization quality. Afterwards, the discriminator is rewarded for correct classifications and the generator for generating examples that fooled the discriminator. Both models are then updated and the next cycle of image normalization begins. 

A triplet loss function is used to train the normalization model $\tau$, including: (a) a reconstruction loss, (b) a GAN loss, and (c) a feature-preserving loss. The objective function $L (\theta)$ can be written as:
\begin{equation}
L(\theta) = \alpha L_{GAN}(\tau, \it{s}^{(i)}) + \beta L_{Recon}(\tau, \it{s}^{(i)}) + \gamma L_{FP}(\zeta, \it{s}^{(i)})
\end{equation}

For the reconstruction loss, we measure the difference between generated images $\dot X_A$ and its original counterparts $X_A$, and enforce the generator to learn an image color distribution and maintain the structural information in images at the same time. Specifically, structural information refers to the knowledge about the structure of objects, e.g. spatially proximate, in the visual scene. In the context of computational histopathology, structural information mainly refers to the spatial organization of histological substances, i.e. multicellular structures, in histopathology images. Such information is key for downstream computational histopathology and thus should be maintained in color normalization. In prior works, generative networks usually adopt mean squared error (MSE) as an image reconstruction loss function. However, MSE-driven models are prone to generating a smoothed/blurred reconstruction where some structural information in the original signal is missing \cite{liang2020stain}. To address this problem, we introduce a loss function based on structural similarity index measure (SSIM) to measure the quality of generated images \cite{wang2004image}. The motivation behind this is that structural similarity correlates well with humans perception of image quality and facilitates the networks to maintain the texture and structural patterns in images \cite{bakurov2022structural}.\\

The SSIM-based reconstruction loss function when optimizing the model can be formulated as:

\begin{equation}
L_{recon}(G) = E_{X\sim X_A}[1-\operatorname{SSIM}(X, G(X, \theta_G))]
\end{equation}

where $L_{recon}(G) \in [0,1]$ and SSIM$(X, G(X, \theta_G))$ is the structural similarity index matrix between original image $X$ and generated image $\dot X$ obtained using the model $G$ that has $\theta_G$ as parameters, which are learned during the training process. As SSIM is proposed for gray-scale images, in practice, we first map RGB images to gray-scale images. A sliding window is applied to obtain gray-scale images and image differences within the sliding windows, characterized by luminance, structure, and contrast, are evaluated and averaged for a single SSIM value.

On the other hand, by using the Deep Convolutional Generative Adversarial Networks architecture (DCGAN), the generator ${\zeta}$ learns a mapping linking $\it{G}$ to $\it{C}$ and also deceives the discriminator $D$, allowing to distinguish between fake and real images. Therefore, we use the following GAN loss:

\begin{equation}
L_{GAN}(\zeta,D) = \mathbb{E}\left[\ln D(X)\right] + \mathbb{E}\left[\ln (1-D(\zeta(G(X))))\right]
\end{equation}

Here, while $D$ learns to maximize $L_{GAN}$, ${\zeta}$ attempts to minimize it until both reach an optimal state. To this point, every stained image might be transferred to have the desired stain style. However, this approach often tends to make frequent color images independent from histological features and this is the reason we introduced an additional loss function, the feature preserving loss that can be formulated as:

\begin{equation}
L_{FP}(\zeta, \hat{f}^{(i)}) = \mathbb{KL}[F(\hat{f}^{(i)}(c)) | F(\hat{f}^{(i)}(\zeta(\dot X_A)))]
\end{equation}

where $L_{FP}(G,D)$ represents the feature preserving loss,  $\zeta$ is the generator and $D$ is the discriminator. Here, $\mathbb{KL}$ denotes the Kullback-Leibler divergence. $F(\hat{f}^{(i)}(.))$ indicates the feature of a given color image extracted from the classifier $\hat{f}^{(i)}$. Here, $f: X\rightarrow Y$ is the tumor classifier network which infers histological patterns from input image $x \in X$, and $c$ is the class of stained images or color images with RGB channels, defined by the set of $d\times{}d$ matrix with $\mathbb{R}^3$ entries and denoted by $C = M_d(\mathbb{R}^3)$. The tumor classifier was trained to distinguish the tumoral nodule from the rest of the tissue. We used 200 binary masks of the tumor core provided by a previous study on part of the same dataset \cite{elforaici2022semi}. By incorporating the feature preserving loss into the training process, we encourage the generator to preserve the important features present in the color space denoted $s^{i}$ on Figure \ref{fig2}.

It is imperative to apply a normalization step at this stage due to the substantial color variations between the histology slides used in the framework, caused by the variations in tissue preparation and the staining procedures. Variations in colors may affect the robustness of the proposed prognostic pipeline and subsequently, its performance.

\subsection{Semi-supervised ViT knowledge distillation network for CLM prognosis}
\subsubsection{Semi-supervised tissue classification}
After normalizing the WSI, we use a semi-supervised approach to classify tissues on the slides into 4 different classes: normal tissue, fibrosis, cancer, and necrosis. It will also provide segmentation of the tumor core and extract features that will be required for the prognosis model.

The proposed semi-supervised learning framework combines annotated and unannotated data to improve model performance. The model learns to make predictions on annotated data and then applies the patterns learned from that data to make predictions on unannotated data. This method is especially beneficial in cases where obtaining annotated data is difficult or expensive, such as in a histopathology context, as it can help to increase model accuracy and generalization without requiring as much annotated data as traditional supervised learning would \cite{hady2013semi}.

\begin{figure*}[t]
    \centering
    \includegraphics[width=18cm]{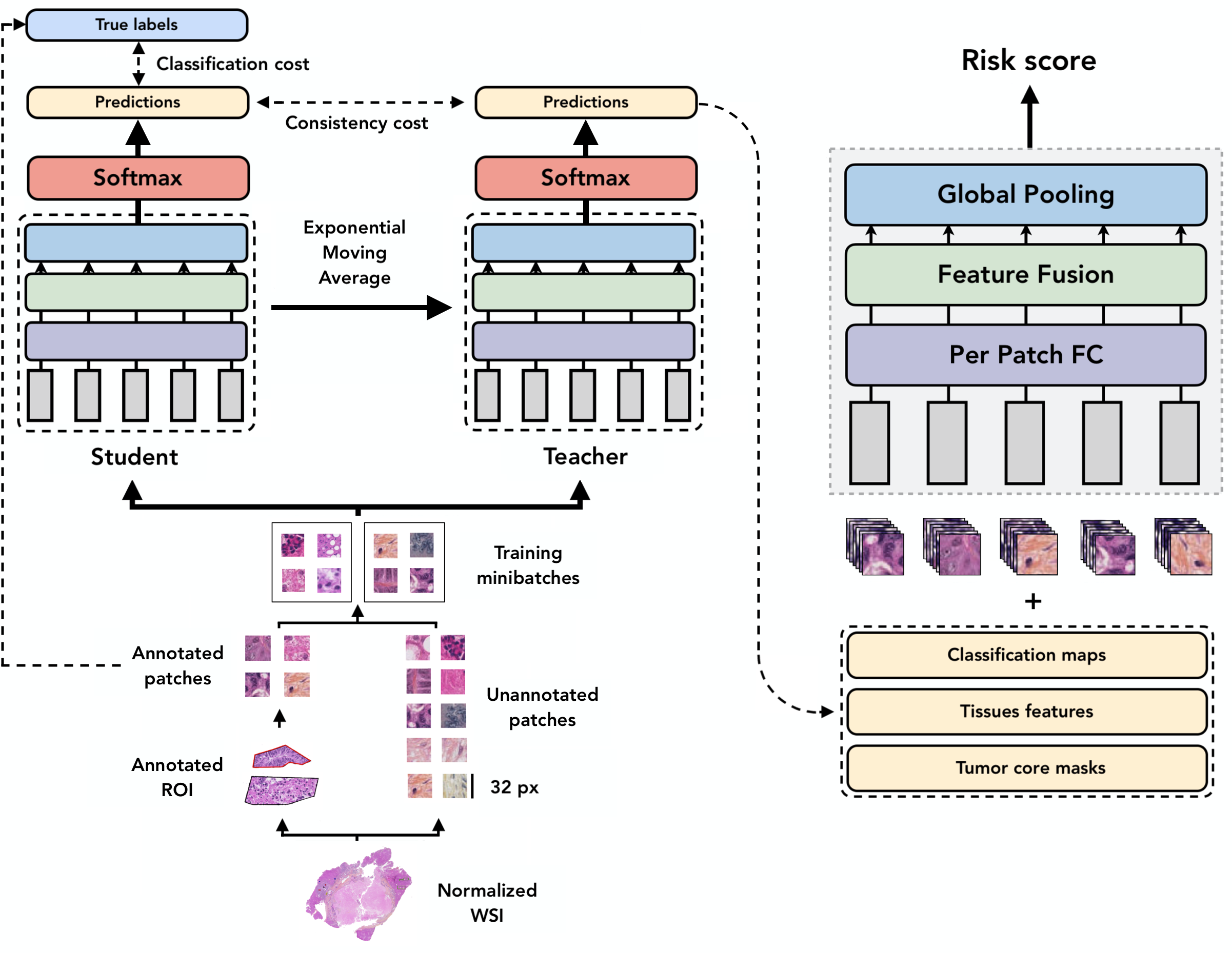}
    \caption[fig3]
    {\label{fig3} Illustration of the Mean Teacher Approach for tissue classification in the context of survival analysis on input patches from histopathology slides. The student and teacher models are jointly trained using exponential moving average (EMA) to generate $p_1$ and $p_2$ tissue class probability distributions, respectively. The loss function is defined as the cross-entropy between both predictions, promoting consistency between the student and teacher models. The teacher model consists of per-patch fully connected layers, followed by feature fusion and global pooling to obtain a risk score. As an output, the model produces risk scores related to survival and TRG of the CLM lesion.}
\end{figure*}

Mean Teacher (MT) model is a semi-supervised approach first proposed by Tarvainen and Valpola \cite{tarvainen2017mean} consisting of two models, a student, and a teacher model, both sharing the same CNN architecture, as shown in Figure \ref{fig3}. Alternatingly, the student and teacher models are updated. The student model (with weights ${\theta}$ and noise ${\eta}$) learns from the teacher model (with weights ${\theta'}$ and noise ${\eta'}$) at each step by minimizing the weighted sum of the loss obtained from annotated data $x$ and the consistency loss \textit{J} obtained from unannotated data. It is defined as the expected distance between the predictions of the student and teacher models, given by:
\begin{equation}
J(\theta)=\mathbb{E}_{x,\eta',\eta}
\Big[|f(x,\theta',\eta')-f(x,\theta,\eta)|^2 \Big]
\end{equation}

The teacher model updates its weights ${\theta'}$ using an exponential moving average (EMA) of the student weights ${\theta}$, as shown by the equation:

\begin{equation}
\theta't=\delta \theta'{t-1} + (1-\delta)\theta_t
\end{equation}

where ${\delta}$ represents the smoothing coefficient hyperparameter.

Both the student and teacher models evaluate the input by incorporating noise into their computation, ${\eta}$ and ${\eta'}$ respectively. The output from the softmax layer of the student model is compared to the one-hot label using a cross-entropy loss (classification cost), and also with the teacher output using the expected distance between the two predictions (consistency cost). Both model outputs can be used for prediction, but it is common to use the teacher’s prediction as it is more likely to yield improved performance compared to the student model. With an unannotated example, the training step is similar, but this time no classification cost is applied.

We trained this model for multiclass classification using a meticulously balanced training set that comprised both annotated and unannotated patches extracted from normalized slides used as inputs. The annotated patches had one of the following four labels: normal tissue, fibrosis, cancer, necrosis, or background. To ensure the reliability and effectiveness of our model, the patches used for training and testing were derived from distinct cohorts of patients, allowing the network to effectively learn from diverse and independent data sources. As an output, our model not only generates highly informative classification maps for each slide based on probability $p_1$ and $p_2$ from the student and teacher models, respectively, where different tissue classes are assigned distinct colors but also provides valuable feature vectors specific to each class that will be the inputs for the next model.

The feature vector output was expressed using the following formula, which represented the output of the model's classification layer for a given input patch. Denoting the feature vector as ${\mathbf{v}}$, it was obtained by feeding the input patch through the MT model and extracting the activations from the desired layer. Let's assume the desired layer is denoted as $L$, which could be the last fully connected layer or a pooling layer.

Mathematically, the feature vector ${\mathbf{v}}$ was obtained as follows:

\begin{equation}
{\mathbf{v}} = f_{L}(x)
\end{equation}

In this equation, $f_{L}(\cdot)$ represents the function that computed the activations at layer $L$ given the input patch ${x}$. During inference, the pre-trained MT model was used by providing the input patch ${x}$, and the activations from the desired layer $L$ were extracted to form the feature vector ${\mathbf{v}}$. This feature vector represented the high-level features specific to one of the four classes.

\subsubsection{Survival / time-to-recurrence and TRG prediction}
The final module of the prognosis framework consists of a combination of multiple layers of fully convolutional layers and non-linear activation functions, which was proven to be a powerful non-linear feature mapping in multiple instance problems \cite{yang2017miml}. The rationale to use fully convolutional networks (FCN) without including any fully connected layers is that FCN is more flexible and can handle any spatial resolution, which is needed for the problem at hand since the number of sample patches in each image varies. For each image, the input is a set of features from $n$ patches, which can be organized as $1 \times n \times d$ ($d$ is the feature dimension or channel). The network consists of several layer-pairs of $1 \times 1$ Conv layer and ReLU layer. The global pooling layer (e.g. average pooling) is added at the end. For $j_{th}$ image, its representation is denoted as $r_j$. The network receives an image as input and it can focus on local information and generate its representation. Hence due to the variations in content and number of patches, the FCN is more flexible to handle this scenario. Two loss functions were used in the prognosis model:
\begin{enumerate}
\item Cox partial likelihood: The first loss function is based on the Cox partial likelihood, which is used for fitting Cox proportional hazard models, and can be extended to train neural networks based on the following term:
\begin{equation}
L_{cpl}=\sum_{i}{ln\left(\sum_{i \in R_i} \exp(f(X_j)-f(X_i))\right)}
\end{equation}

where $R_i$ includes individuals with event times at $T_i$ that is the event time or time of last follow-up, $X_i$ is the set of whole slide images, and $f(X_i)$ is the risk score, each for the $i^{th}$ case.

\item Censored Cross-Entropy: It is an extension of the standard cross-entropy loss used for classification models, adapted to train survival prediction models using right-censored data. Censorship occurs when the event of interest (e.g., recurrence or death) has not occurred during the study period or follow-up, resulting in incomplete information. To address this, we modeled survival prediction as a classification problem by discretizing time into intervals and training models to predict the discrete time interval in which the event occurred. The loss is defined as follows:

\begin{equation}
L_{cc}={} \sum\limits_{i} \left[
\begin{aligned}
& O_i \ln(f(X_i)[Y_i]) \\
& + (1-O_i) \ln(\sum_{y>Z_i}{f(X_i)[y]}) \\
\end{aligned}
\right]
\end{equation}

where $Y_i$ is the interval in which the event occurred (for example with observed events) and $Z_i$ is the latest interval whose endpoint is before the time of censorship. Here, $f(x)$ is a predicted probability distribution over time intervals, and $f(X_i,y)$ is the probability assigned by the model for the event occurring in the $y^{th}$ interval. To obtain a scalar risk score for evaluation, we took the negative of the expectation over the predicted time interval likelihood distribution. The negation ensures that higher risk score values indicate higher risk.

\end{enumerate}
Once the proposed model and the KD models are trained, we use the generated classification maps after average pooling from the classification step for the TRG prediction. Using the normal tissue distribution, we extracted the metastatic nodules from the images. These nodules and their surrounding tissue contain all the necessary information for predicting the TRG. We combine the visual features with the TRG scores from our clinical dataset to train a CNN for TRG prediction. In this work, we experimented classification with 2 and 3 classes.

\subsubsection{Knowledge distillation}
In order to improve performance on the survival and time-to-recurrence prediction task, knowledge is transferred from our semi-supervised Mean Teacher model to a ViT which leverages the strengths of both models. Our semi-supervised pipeline is based on a CNN architecture that is known to be very effective at learning local, spatial features in images \cite{alzubaidi2021review}, while ViTs are better suited for capturing long-range dependencies and global context \cite{hatamizadeh2022global}. By combining the strengths of both models, we can potentially achieve better performance than either model on its own \cite{yang2022vitkd}.

Adversarial distillation based on GANs, is one of the techniques that was proposed to improve the process of transferring knowledge \cite{gou2021knowledge}. Similarly to the normalization task that was accomplished using a GAN, this task will train a discriminator to estimate the probability that a sample comes from the training data distribution while the generator tries to deceive the discriminator using generated data samples. In this case, we  use a teacher model as a discriminator and a student model as a generator. For a certain risk score, the student model will provide histology slide patches to deceive the teacher model. After the training phase, the student model is able to deceive the discriminator, and the quality of the generated patches matches the real patches. The distillation loss used in this GAN can be formulated as:

\begin{equation}
\begin{split}
L_{GAN-KD} = & L_{CE}(G(F_s(x)), y) + \alpha_1 L_{KL}(G(F_s(x)), F_t(x))\\ 
& + \alpha_2 L_{GAN}(F_s(x), F_t(x))
\end{split}
\end{equation}

where $L_{CE}$ is the cross-entropy loss, $L_{KL}$ is the Kullback-Leibler (KL) divergence loss, commonly used in generative adversarial networks. $F_t$ and $F_s$ are the outputs of the teacher and student models, respectively. $G(F_s(x))$ indicates the training samples generated by the generator $G$. Finally, $\alpha_1$ and $\alpha_2$ are hyperparameters.

We considered four different distilled models, and trained them using contrastive representation distillation (CRD) \cite{tian2019contrastive} with our SSL model. The first model is a SSL model pre-trained on tissue classification and TRG prediction \cite{elforaici2022semi}. The other three models are ViTs: target-aware transformer (TaT), GasHis-Transformer, and pyramid vision transformer (PVT).

\begin{enumerate}
\item TaT is designed to learn the target-specific knowledge of the teacher model and transfer it to the student model. It is a recent one-to-all spatial matching knowledge distillation approach that allows each pixel of the teacher feature to be distilled to all spatial locations of the student features given its similarity, which is generated from a target-aware transformer \cite{lin2022knowledge}.

\item GasHis-Transformer is a multi-scale visual transformer approach for gastric histopathological image detection. The idea of the multi-scale architecture is introduced to describe the details of gastric tissues and cells under a microscope. GasHis-Transformer not only obtains good classification performance on gastric histopathological images but also shows an excellent generalization ability on other histopathological image datasets. The model consists of two key modules designed to extract global and local information using a position-encoded transformer model and a convolutional neural network with local convolution, respectively \cite{chen2022gashis}.

\item PVT is the first pure transformer backbone designed for various pixel-level dense prediction tasks. It is composed of a stack of multi-scale feature maps generated by a pyramidal structure. Each feature map is processed by a transformer encoder to capture global context information and a CNN to extract local features. The overlapping patch embedding design is used to reduce the computational complexity of the transformer encoder. The linear complexity attention layer design is used to improve the efficiency of the transformer encoder. The convolutional feed-forward network design is used to enhance the local feature extraction ability of the CNN \cite{wang2021pyramid}.
\end{enumerate}

\subsection{Experimental setup}
\subsubsection{Normalization}
We use stochatstic gradient descent (SGD) to optimize the diagnosis network with a learning rate of $10^{-3}$, and a batch size of 8 on the training set for 100 epochs. SGD was then used with a learning rate of $10^{-4}$ and a batch size of 4 to train the GAN-based style transfer model on the training set for 60 epochs. Hyper-parameters
(i.e. weights of different loss functions) are tuned using the validation set. We finally chose $\alpha = 0.2$, $\beta = 0.3$, $\gamma = 0.5$.
To evaluate the proposed method, we tested the learning efficiency of our model and the effectiveness of the SSIM-based loss function in GAN training. To this end, we recorded the image reconstruction loss in training to trace the optimization procedure.

\subsubsection{Classification}
Using the baseline annotated training set, the remaining patches were used as the unannotated dataset for semi-supervised learning (SSL). According to Yalniz et al. (2019), the pseudo-annotated dataset can be exploited in two ways, using the $P$ and $K$ parameters \cite{yalniz2019billion}:
\begin{enumerate}
\item For each training instance, only the top $P$ class probabilities are kept as non-zero.
\item Only the top $K$ training instances are retained for each class, where instances are classified according to the highest probability class and the highest probability instances are retained.
\end{enumerate}
Optimization of hyperparameters $K$ and $P$ was performed on a single iteration (first student) of the teacher-student semi-supervised loop, using  a portion of the annotated data, and the validation set from the same dataset. We found that the application of the $P$ parameter had little impact and therefore we chose to retain all class labels. However, the use of the $K$ parameter (i.e. only retaining $K$ patches with the highest confidence in each class) provided a modest improvement in performance with the optimal value being $K$ = 4000 per class in the pseudo-annotated training set. Here, $K$ corresponds to approximately 80\% of the available unannotated training set, i.e. a small proportion of the least certain examples for each class is discarded. The annotated patches were randomly distributed as follows: 70\% for the training set, 15\% for validation and 15\% for testing. During training, both networks use SGD with momentum as the optimizer to alternately train the target model for 100 epochs. The momentum rate is set to 0.9. The learning rate is 0.001, and the batch size of the labeled and unannotated data sets is 32.
Student t-tests were performed to evaluate statistical significance. A two-sided p-value < 0.05 was considered statistically significant. Statistical analyses were conducted using Python Scipy v1.5.4, Python Lifelines v0.27.1, scikit-survival v0.20.0, and R Survival v3.4 packages.

\subsubsection{Tumor aggregation strategies}
It is important to consider different strategies for aggregating features from multiple liver metastases in a patient to train the survival prediction method. Different strategies of tumor aggregation can significantly impact the performance of predictive models \cite{yao2020whole}. In this study, we considered three aggregation strategies: 

\begin{enumerate}
\item Max pooling: Selects the maximum value across all feature maps for each tumor in the separate WSI. This approach emphasizes the most important features across all the slides.

\item Mean pooling: Calculates the average value of all feature maps for each ROI in the different slides. This approach emphasizes the common features across all the slides.

\item Weighted average pooling: Uses the lesion volume to weigh the importance of the features from each lesion. This approach can learn to emphasize the most informative features across all slides and can adapt to different levels of heterogeneity among the slides.

\end{enumerate}

\section{Results}

\subsection{Normalization}
In order to evaluate the performance of our normalization model, we compared it to two other commonly used methods for staining normalization: Macenko \cite{macenko2009method} and Reinhard \cite{reinhard2001color}. As shown in Table \ref{tab1}, we also selected two evaluation techniques that are used to assess the quality of the image normalization process: the Structure Similarity Index Matrix (SSIM) \cite{wang2004image} and the Pearson correlation coefficient (PCC) \cite{rodgers1988thirteen}.

On one hand, SSIM is used to measure the structural information, luminance, and contrast between the source and processed image and its index denotes the reference metric:

\begin{equation}
\operatorname{SSIM}(x,y)=(\frac{2\mu_x\mu_y+\ c_1}{\mu_x^2+\mu_y^2+\ c_1})\ (\frac{2\sigma_{xy}+\ c_2}{\sigma_x^2+\sigma_y^2+\ c_2})\\
\end{equation}

On the other hand, PCC measures the linear correlation between the two images and its range from 0 to 1. A value of 0 indicates that there is no similarity between the two images:

\begin{equation}
\operatorname{PCC}(x,y)=\ \frac{\sum_{i}\left(x_i-\ \mu_x\right)\left(y_i-\ \mu_y\right)}{\sqrt{\sum_{i}\left(x_i-\ \mu_x\right)^2}\sqrt{\sum_{i}\left(y_i-\ \mu_y\right)^2}}
\end{equation}

\begin{table}
\caption {Normalization performance of the proposed model compared to Macenko and Reinhard algorithms using the SSIM and PCC evaluation measures.}
\centering
\scalebox{0.9}{
\begin{tabular}{|c|c|c|}
\hline
\multirow{2}{*}{\textbf{Normalization method}} & \multicolumn{2}{c}{\textbf{Evaluation algorithm}} \vline  \\ \cline{2-3} 
& \textbf{SSIM} & \textbf{PCC} \\ 
\hline
\hline
\textit{Macenko \cite{macenko2009method}} & 0.864 & 0.831 \\
\hline
\textit{Reinhard \cite{reinhard2001color}} & 0.898 & 0.852 \\
\hline
\textit{Our GAN model} & \textbf{0.913} & \textbf{0.887} \\
\hline
\end{tabular}
\renewcommand{\arraystretch}{1}
\label{tab1}
}
\end{table} 

\begin{figure*}[t]
    \centering
    \includegraphics[width=18.5cm]{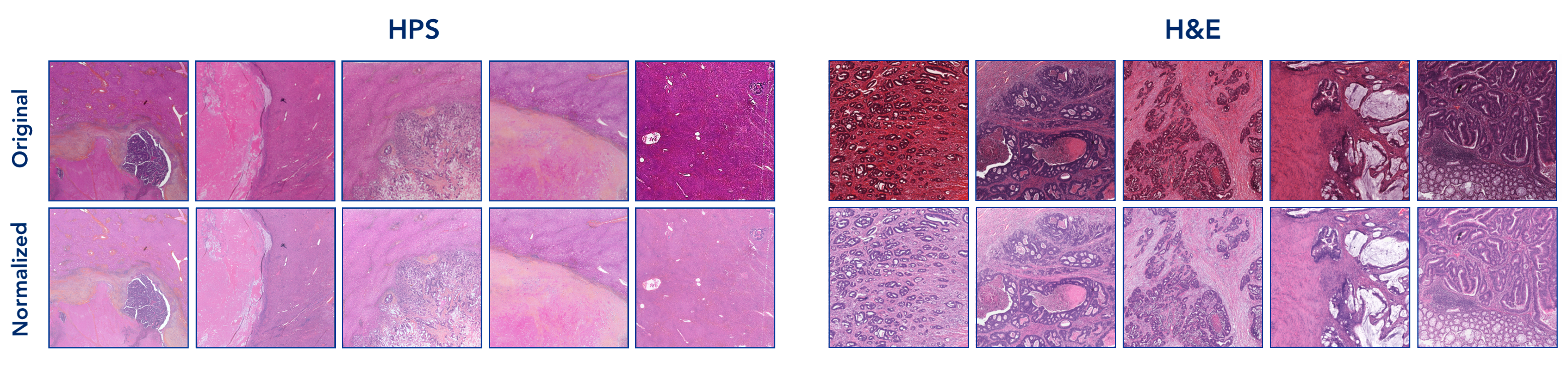}
    \caption[fig4]
    {\label{fig4} Sample results of normalized slides with the proposed GAN model. We can see a homogeneity in the color distribution in the obtained slides despite variability in the original slides, that can be either HPS or H\&E stained.}
\end{figure*}

The proposed model yielded significant improvement in both metrics. Sample results of the normalized slide obtained with the GAN model are shown in Figure \ref{fig4}.

\subsection{Tissue classification}

\begin{figure*}[!t]
    \centering
    \includegraphics[width=18cm]{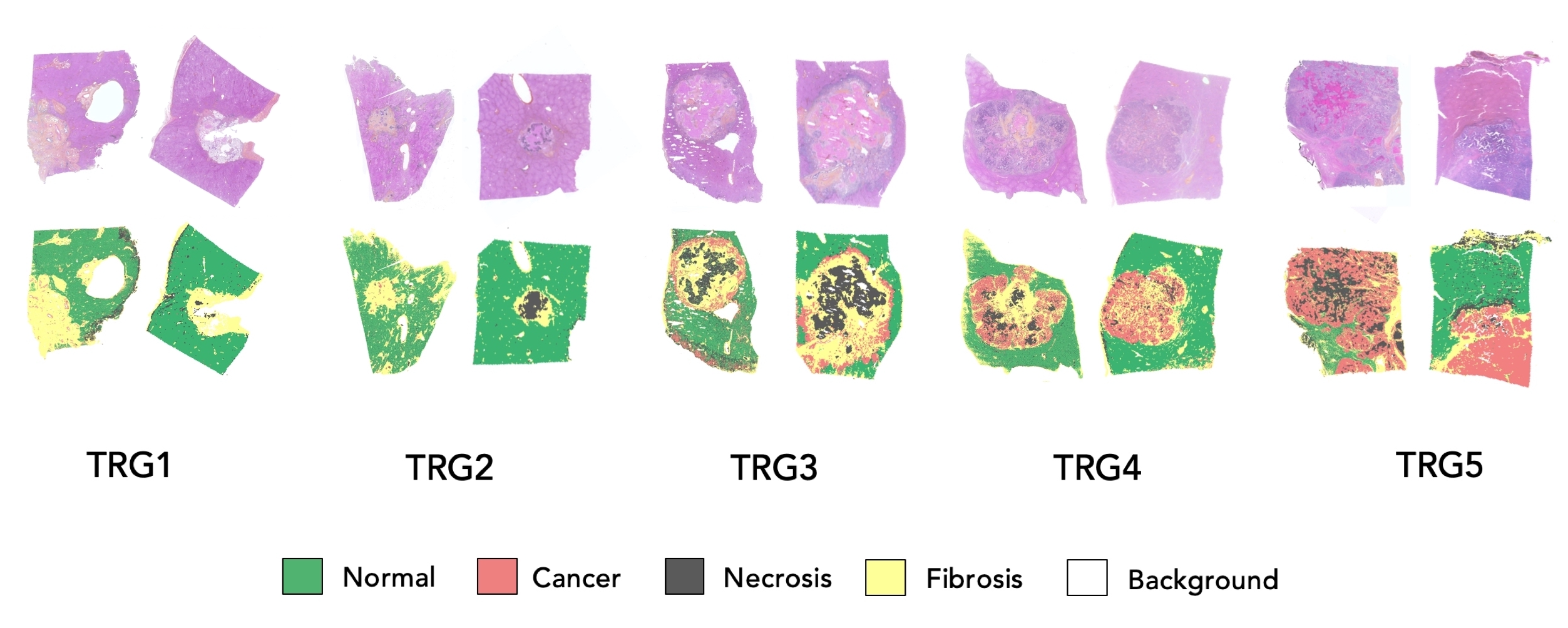}
    \caption[fig5]
    {\label{fig5} Sample results for the classification task. In the first row, we show two normalized slides for each TRG score. In the second row, the correspondent classification map is generated with the SST model.}
\end{figure*}

Figure \ref{fig5} shows some examples of the classification maps obtained using the normalized slides as input. This illustrates the performance of our model in the tissue classification task, where each 32 x 32 \textit{px} patch from the WSI is represented by one pixel on the classification map.

We can observe the evolution of the classification maps with regard to the TRG score. In the case of the TRG 1 for instance, the slides are completely covered in yellow and green segmentation, which indicates an absence of residual cancer and large amounts of fibrosis. This suggests that the neoadjuvant therapy was effective in reducing the size of the tumor, and the model is able to accurately classify the different tissue types in the slide. Meanwhile in the case of TRG5, the tumor core is mostly red which indicated a predominance of tumor cells along with small necrotic areas in black, suggesting that the neoadjuvant therapy was not effective in reducing the size of the tumor.

The proposed classification model is compared to three comparative methods, namely VAT \cite{miyato2018virtual} and TSchain \cite{shaw2020teacher}, which are both semi-supervised models, as well as the ResNet-50 supervised model.

Table \ref{tab2} summarizes the performances with regards to the accuracy and F1-score. Our model outperforms all other models on the test set composed of normalized H\&E slides (statistically significant). The relatively weak performance of the supervised model can be explained by the few annotations available in our dataset that give an advantage to semi-supervised approaches.

\renewcommand{\arraystretch}{1}
\begin{table}[h]
\caption {Tissue classification task performances. The proposed model is compared to two semi-supervised models (VAT and TSchain) and a supervised model (ResNet-50). Statistically significant results shown in bold.}
\centering
\scalebox{0.9}{
\begin{tabular}{|c|c|c|}
\hline
\multirow{2}{*}{\textbf{Models}} & \multicolumn{2}{c}{\textbf{Classification performance}} \vline  \\ \cline{2-3} 
& \textbf{Accuracy} & \textbf{F1 score} \\ 
\hline
\hline
\textit{ResNet-50} & 80.3\% (0.7) & 79.8\% \\
\hline
\textit{VAT \cite{miyato2018virtual}} & 87.5\% (0.8) & 88.2\% \\
\hline
\textit{TSchain \cite{shaw2020teacher}} & 91.6\% (0.7) & 90.5\% \\
\hline
\textit{Proposed SSL model} & \textbf{94.8}\% (0.7) & \textbf{92.3\%} \\
\hline
\end{tabular}
\renewcommand{\arraystretch}{1}
\label{tab2}
}
\end{table}

\subsection{Prognosis prediction}
The following set of experiments evaluated the prediction of OS and TTR. Three comparative methods were used in this work: (a) MobileNetV2 \cite{sandler2018mobilenetv2}, (b) MobileNetV3 \cite{howard2019searching}, and (c) Deep Attention Multiple Instance Survival Learning (DeepAttnMISL) \cite{yao2020whole}.

\begin{figure}[h]
    \centering
    \includegraphics[width=9cm]{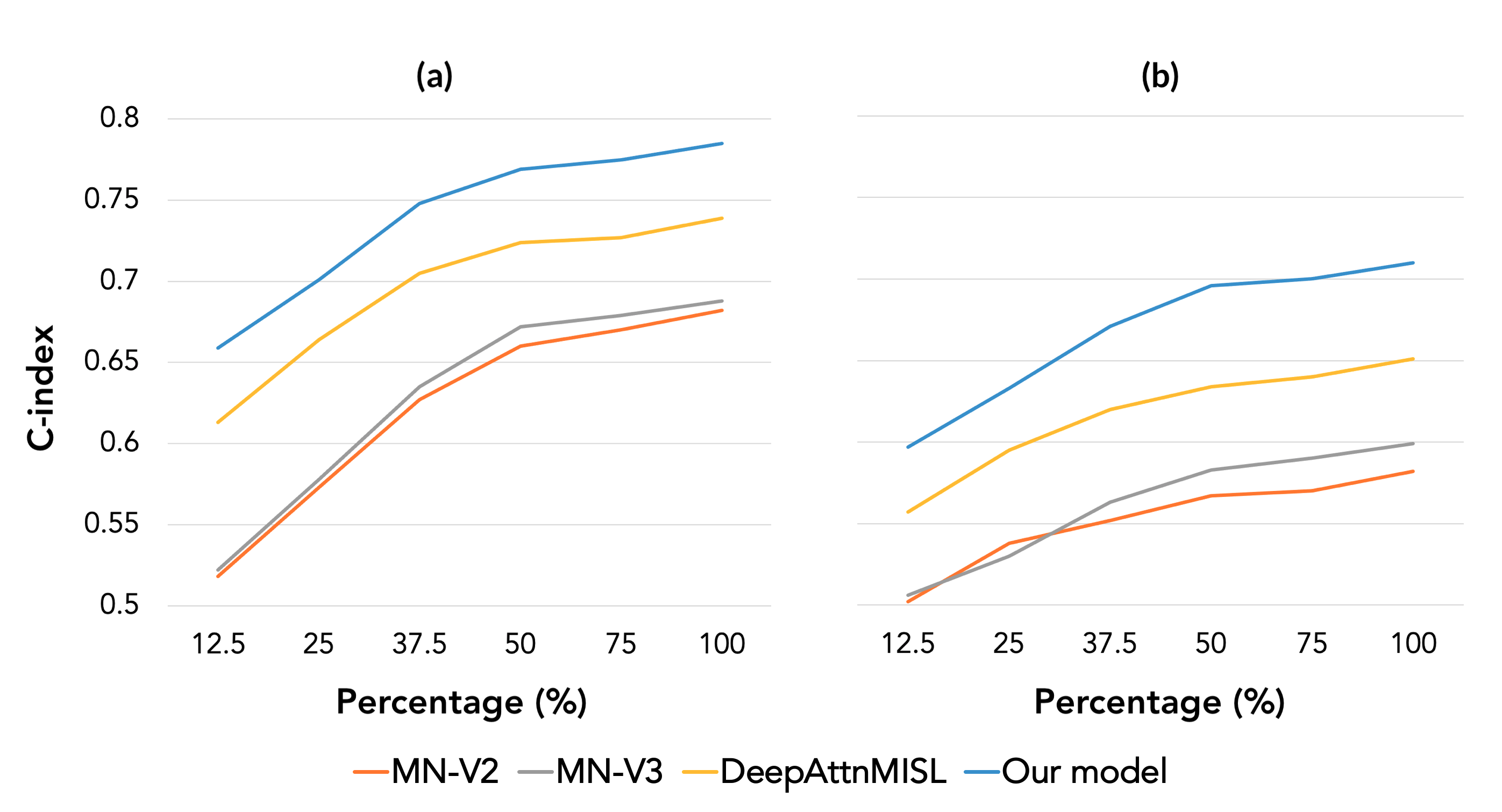}
    \caption[fig6]
    {\label{fig6} Model performances for survival and TTR. It is represented in c-index curves for (a) OS prediction, and (b) TTR. We can observe that the proposed model yields improved accuracy to the comparative models for both prediction tasks.}
\end{figure}

In order to compare the performances of the models, we show in Figure \ref{fig6} the concordance index, also known as the c-index, obtained by each model in six scenarios, using 12.5\%, 25\%, 37.5\%, 50\%, 75\% and 100\% of labeled data. The c-index is a widely-used metric that quantifies the discriminatory power of survival models. It measures the proportion of concordant pairs, where cases are appropriately ordered based on their predicted risk scores, taking into account both the observed events and the censored cases. In this context, informative pairs are pairs of cases with distinct outcomes and different censorship times, which provide valuable information for evaluating the performance of the models. By considering these informative pairs, the concordance index captures the ability of the model to accurately rank the cases and reflects its discriminative performance in survival prediction tasks. Figure \ref{fig6} illustrates the obtained concordance index for each model in the six scenarios, shedding light on their comparative performance.

\begin{figure*}[!t]
    \centering
    \includegraphics[width=18cm, height=7.75cm]{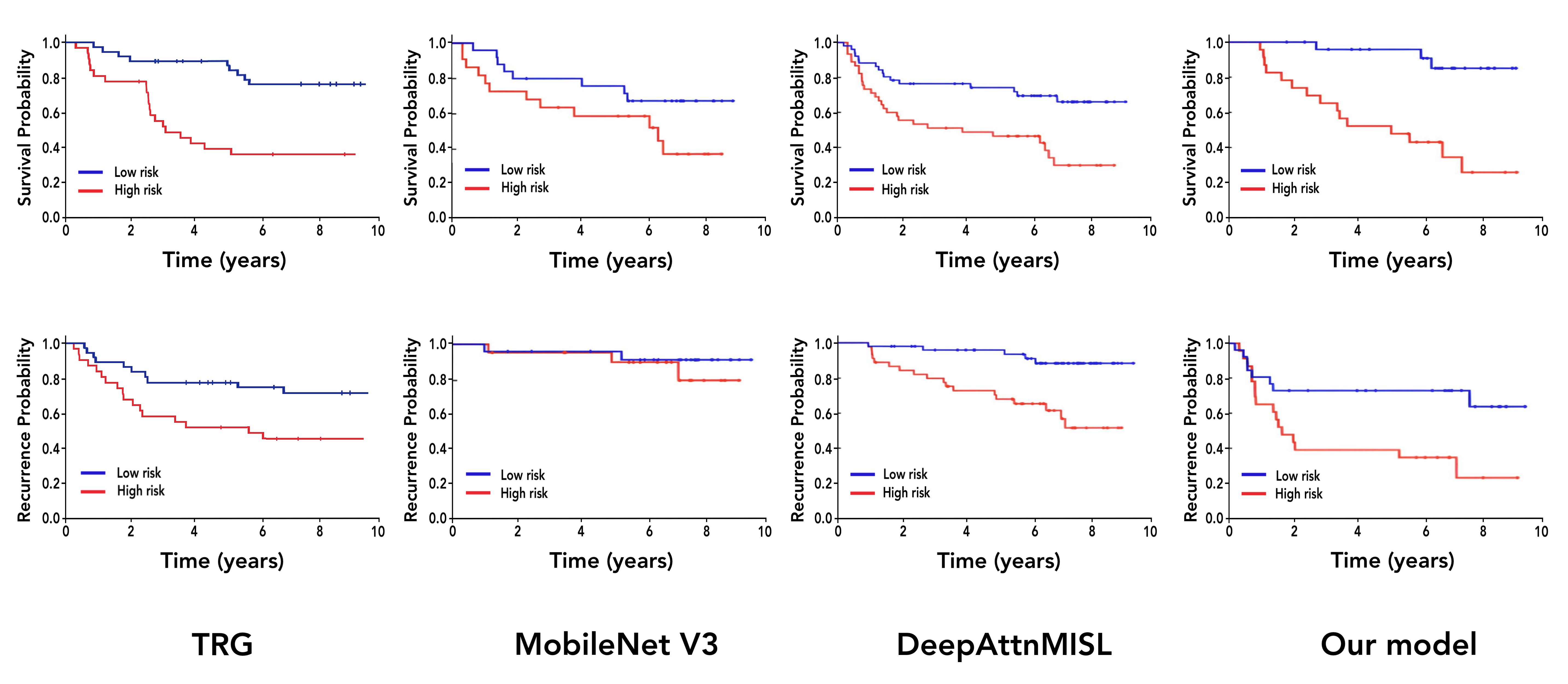}
    \caption[fig7]
    {\label{fig7} Kaplan-Meier curves for OS prediction (first row) and TTR (second row). From left to right: TRG stratification (1-2 vs 3-5), MobileNetV3, DeepAttnMISL, and our model (SSL + KD).}
\end{figure*}

In the OS prediction task, our approach achieved the highest performance, which was statistically better than the DeepAttnMISL approach. On the other hand, the MobileNet models (V2 and V3) yielded an average accuracy 9\% lower than the proposed approach and 5\% lower than DeepAttnMISL method. In previous work \cite{elforaici2022semi}, we demonstrated that the limited quantity of annotated data used for training provided an advantage to the semi-supervised approaches over the supervised one that uses essentially annotated patches to train. In a clinical setting, the scarcity of annotations is frequently observed.
Using TRG data and three trained models (MN-V3, DeepAttnMISL, and our model), we generated Kaplan-Meier curves shown in Figure \ref{fig7}. These curves demonstrate a significant risk stratification with our method, where the distinction between low and high risk patients if significantly larger for both OS and TTR as opposed to the comparative models.

\subsection{TRG prediction}

We evaluated the performance of our proposed approach for predicting the Tumor Regression Grade (TRG) in comparison to the DeepAttnMISL model. The following two tables summarize accuracy results for the tested models applied to different TRG classifications, in two and three classes. 
\\
Our approach achieved higher accuracy than the DeepAttnMISL model across all dichotomizations and classifications. For the dichotomizations, our approach outperformed DeepAttnMISL with statistically significant improvements in accuracy, as shown in Table \ref{tab3}. In the classifications, our approach consistently achieved higher accuracy compared to DeepAttnMISL, with statistically significant performance differences, as shown in Table \ref{tab4}.

\renewcommand{\arraystretch}{1}
\begin{table}[h]
\caption {TRG prediction performance for our approach, compared to DeepAttnMISL model, in 3 different TRG dichotomizations. Statistically significant results are shown in bold.}
\centering
\scalebox{0.9}{
\begin{tabular}{|c|c|c|}
\hline
\multirow{2}{*}{\textbf{TRG}} & \multicolumn{2}{c}{\textbf{Classification performance}} \vline  \\ \cline{2-3} 
& \textbf{DeepAttnMISL} & \textbf{Our model} \\ 
\hline
\hline
\textit{1 vs 2-5} & 83.8\% (1.4) & \textbf{86.9\% (1.5)}\\
\hline
\textit{1-2 vs 3-5} & 88.4\% (1.2)& \textbf{90.3\% (1.4)}\\
\hline
\textit{1-3 vs 4-5} & 84.5\% (1.6) & \textbf{87.3\% (1.7)}\\
\hline
\end{tabular}
\renewcommand{\arraystretch}{1}
\label{tab3}
}
\end{table} 

\renewcommand{\arraystretch}{1}
\begin{table}[h]
\caption {TRG prediction performance for our approach, compared to DeepAttnMISL model, in 3 different TRG classifications. Statistically significant results shown in bold.}
\centering
\scalebox{0.9}{
\begin{tabular}{|c|c|c|}
\hline
\multirow{2}{*}{\textbf{TRG}} & \multicolumn{2}{c}{\textbf{Classification performance}} \vline  \\ \cline{2-3} 
& \textbf{DeepAttnMISL} & \textbf{Our model} \\ 
\hline
\hline
\textit{1 vs 2-3 vs 4-5} & 78.2\% (1.7) & \textbf{81.7\% (1.8)} \\
\hline
\textit{1-2 vs 3 vs 4-5} & 78.3\% (1.6) & \textbf{82.1\% (1.8)} \\
\hline
\textit{1-2 vs 3-4 vs 5} & 72.4\% (1.8) & \textbf{78.5\% (1.9)} \\
\hline
\end{tabular}
\renewcommand{\arraystretch}{1}
\label{tab4}
}
\end{table}

\subsection{Ablation experiments}
Here, we present the results of ablation studies conducted to evaluate the impact of different loss functions on WSI normalization. The evaluation was performed using two metrics: SSIM and PCC.
The results, shown in Table \ref{tab5}, indicate that the combination of $L_{GAN}$, $L_{Recon}$, and $L_{FP}$ loss functions achieved the highest performance in terms of both SSIM and PCC. These results highlight the importance of employing multiple loss functions for WSI normalization, leading to improved performance.
Among the individual loss functions, $L_{Recon}$ resulted in the highest SSIM, while $L_{GAN}$ yielded the highest PCC. Combining $L_{GAN}$ with $L_{Recon}$ significantly improved both SSIM and PCC scores. However, when combined with $L_{FP}$ individually, the performance was slightly lower.

Table \ref{tab6} compares the performance of different models with and without the normalization step. The models were evaluated based on the c-index for TTR and OS.
Without normalization, the SSL model achieved moderate c-index values for both TTR and OS. Incorporating knowledge distillation (KD) slightly improved the performance. However, when normalization was applied to the SSL model, a significant performance boost was observed, resulting in higher c-index values for TTR and OS. Combining normalization with KD further improved the performance, achieving the highest c-index values.

\renewcommand{\arraystretch}{1}
\begin{table}[h]
\caption {Ablation studies results for the WSI normalization. Top results are indicated in bold.}
\centering
\scalebox{0.9}{
\begin{tabular}{|c|c|c|}
\hline
\multirow{2}{*}{\textbf{Loss functions}} & \multicolumn{2}{c}{\textbf{Evaluation metric}} \vline  \\ \cline{2-3} 
& \textbf{SSIM} & \textbf{PCC} \\ 
\hline
\hline
\textit{$L_{GAN}$} & 81.0\% & 82.5\% \\
\hline
\textit{$L_{Recon}$} & 83.8\% & 77.1\% \\
\hline
\textit{$L_{FP}$} & 75.0\% & 76.1\% \\
\hline
\textit{$L_{GAN} + L_{Recon}$} & \textbf{87.2}\% & \textbf{84.7}\% \\
\hline
\textit{$L_{GAN} + L_{FP}$} & 83.3\% & 82.8\% \\
\hline
\textit{$L_{Recon} + L_{FP}$} & 85.7\% & 79.8\% \\
\hline
\hline
\textit{$L_{GAN} + L_{Recon} + L_{FP}$} & \textbf{91.3}\% & \textbf{88.7}\% \\
\hline
\end{tabular}
\renewcommand{\arraystretch}{1}
\label{tab5}
}
\end{table}

\begin{table}[h]
\caption{Model performance (c-index) with and without the normalization step.}
\centering
\scalebox{0.9}{
\begin{tabular}{|c|c|c|c|}
\hline
\multirow{3}{*}{\textbf{Model}} & \multirow{3}{*}{\textbf{Normalization}} & \multicolumn{2}{c|}{\textbf{Outcome}} \\ \cline{3-4} 
& & \textbf{TTR} & \textbf{OS} \\ 
\hline
\hline
\textit{\centering SSL model} & \multirow{2}{*}{\centering No} & 0.622 (0.025) & 0.701 (0.027) \\
\cline{1-1}\cline{3-4}
\textit{\centering SSL + KD} & & 0.629 (0.018) & 0.711 (0.017) \\
\hline
\textit{\centering SSL model} & \multirow{2}{*}{\centering Yes} & 0.721 (0.021) & 0.789 (0.023) \\
\cline{1-1}\cline{3-4}
\textit{\centering SSL + KD} & & \textbf{0.733 (0.014)} & \textbf{0.804 (0.015)} \\
\hline
\end{tabular}
}
\renewcommand{\arraystretch}{1}
\label{tab6}
\end{table}

\subsection{Overall survival prediction with knowledge distillation}
Finally, we trained four distilled models with the proposed SSL model on our dataset and compared their respective performances. We focused on the use of ViT models as distilled models, except for the KD-SSL, where we used a semi-supervised model pre-trained on tissue classification and TRG prediction. 
We compared the performance of various models on TTR and OS prediction tasks using different tumor aggregation strategies, namely max pooling, mean pooling, and weighted average pooling, as shown in Table \ref{tab7}. The proposed SSL model and KD-PVT consistently outperformed with statistically significant improvements to the comparative methods across both prediction tasks, as indicated by the higher concordance index scores.
The superior performance of the SSL model and KD-PVT can be attributed to several key factors. First, the SSL model leverages the power of semi-supervised learning, which allows it to learn rich representations from unlabeled data, capturing underlying patterns and structures in the tumor data. This enables the model to better generalize to new instances and make more accurate predictions.
Additionally, KD-PVT incorporates knowledge distillation that helps the student model to learn from the rich insights and generalization capabilities of the larger teacher model from the proposed SSL model. By distilling the knowledge, KD-PVT achieves a good balance between model complexity and performance. Furthermore, KD-PVT benefits from the utilization of the PVT (Pyramid Vision Transformer) architecture, which has demonstrated strong performance in a wide range of computer vision tasks. The hierarchical and multi-scale nature of the PVT architecture allows it to effectively capture both local and global tumor features, leading to more accurate predictions.

\begin{table*}[t]
\caption {Model performances expressed in concordance index comparing different models on TTR and OS prediction using various tumor aggregation strategies: 1. max pooling, 2. mean pooling, and 3. weighted average pooling. Top result is shown in bold.}
\centering
\begin{tabular}{|c|cccccc|}
\hline
\multirow{3}{*}{\textbf{Models}} & \multicolumn{6}{c|}{\textbf{Concordance index}}                                                                                                                                                        \\ \cline{2-7} 
                        & \multicolumn{3}{c|}{\textbf{TTR}}                                                                            & \multicolumn{3}{c|}{\textbf{ST}}                                                        \\ \cline{2-7} 
                        & \multicolumn{1}{c|}{\textbf{1}}    & \multicolumn{1}{c|}{\textbf{2}}  & \multicolumn{1}{c|}{\textbf{3}}             & \multicolumn{1}{c|}{\textbf{1}}    & \multicolumn{1}{c|}{\textbf{2}}  & \textbf{3}             \\ \hline
{MN-V2}          & \multicolumn{1}{c|}{0.582 (0.036)} & \multicolumn{1}{c|}{0.588 (0.034)} & \multicolumn{1}{c|}{0.595 (0.031)} & \multicolumn{1}{c|}{0.682 (0.033)} & \multicolumn{1}{c|}{0.685 (0.037)} & 0.679 (0.034) \\ \hline
{MN-V3}          & \multicolumn{1}{c|}{0.599 (0.033)} & \multicolumn{1}{c|}{0.606 (0.034)} & \multicolumn{1}{c|}{0.607 (0.030)} & \multicolumn{1}{c|}{0.688 (0.034)} & \multicolumn{1}{c|}{0.689 (0.035)} & 0.682 (0.035) \\ \hline
{DeepAttnMISL}   & \multicolumn{1}{c|}{0.651 (0.026)} & \multicolumn{1}{c|}{0.662 (0.024)} & \multicolumn{1}{c|}{0.664 (0.023)} & \multicolumn{1}{c|}{0.739 (0.023)} & \multicolumn{1}{c|}{0.742 (0.025)} & 0.737 (0.025) \\   \hline
\textbf{Proposed SSL model}      & \multicolumn{1}{c|}{\textbf{0.710 (0.023)}} & \multicolumn{1}{c|}{\textbf{0.719 (0.023)}} & \multicolumn{1}{c|}{\textbf{0.721 (0.021)}} & \multicolumn{1}{c|}{\textbf{0.785 (0.022)}} & \multicolumn{1}{c|}{\textbf{0.789 (0.023)}} & \textbf{0.786 (0.025)} \\
\hline \hline
{KD – SSL}    & \multicolumn{1}{c|}{0.705 (0.021)} & \multicolumn{1}{c|}{0.712 (0.022)} & \multicolumn{1}{c|}{0.715 (0.021)} & \multicolumn{1}{c|}{0.789 (0.021)} & \multicolumn{1}{c|}{0.791 (0.023)} & 0.789 (0.023) \\ \hline
{KD – TaT}       & \multicolumn{1}{c|}{0.684 (0.019)} & \multicolumn{1}{c|}{0.688 (0.018)} & \multicolumn{1}{c|}{0.688 (0.017)} & \multicolumn{1}{c|}{0.753 (0.016)} & \multicolumn{1}{c|}{0.753 (0.016)} & 0.750 (0.017) \\ \hline
{KD – GasHis}    & \multicolumn{1}{c|}{0.695 (0.017)} & \multicolumn{1}{c|}{0.669 (0.015)} & \multicolumn{1}{c|}{0.701 (0.015)} & \multicolumn{1}{c|}{0.742 (0.014)} & \multicolumn{1}{c|}{0.744 (0.015)} & 0.742 (0.016) \\ \hline \hline

\hline
\textbf{Proposed KD – PVT}       & \multicolumn{1}{c|}{\textbf{0.728 (0.016)}} & \multicolumn{1}{c|}{\textbf{0.733 (0.016)}} & \multicolumn{1}{c|}{\textbf{0.733 (0.014)}} & \multicolumn{1}{c|}{\textbf{0.801 (0.014)}} & \multicolumn{1}{c|}{\textbf{0.804 (0.015)}} & \textbf{0.802 (0.017)} \\ \hline
\end{tabular}
\label{tab7}
\end{table*}

\section{Discussion}
In this study, we proposed an end-to-end approach for the prognosis prediction of CLM patients. We first trained a GAN model to normalize histology slides stained with H\&E and HPS, which is a time-consuming task. Then, we trained a semi-supervised model to perform tissue classification from sparse annotations, generating segmentation and feature maps using attention mechanisms. Using the features extracted for the metastatic nodules and surroundings, we trained a prognosis model to predict OS, TTR, and TRG. In addition, we used contrastive representation distillation to train a ViT to accomplish the same prognosis prediction task.

We demonstrated the feasibility of a normalization model based on GAN for medical image analysis and evaluated its performance against two commonly used methods, Macenko and Reinhard, based on two evaluation techniques, the SSIM and PCC. Results showed that our proposed model outperformed both methods in terms of SSIM and PCC, achieving a SSIM score of 0.91 and a PCC score of 0.88, indicating that our normalization model is capable of producing high-quality normalized images. These results were further supported by the visual comparison of normalized slides, where our model successfully achieved a homogeneity in color distribution despite variability in the original slides due essentially to the staining techniques (H\&E and HPS).

In addition, we also evaluated the performance of our model in survival prediction tasks, and compared it against three models. Our model outperformed all comparative models, achieving the highest performance in both survival and time-to-recurrence prediction tasks, as demonstrated by the c-index curves. Specifically, our approach achieved an average accuracy that was 9\% higher than that of MobileNetV2 and MobileNetV3, and was comparable to the state-of-the-art Deep Attention Multiple Instance Survival Learning approach. These results suggest that our proposed method is a promising approach for survival prediction of CLM treated with chemotherapy.

One possible explanation for the superior performance of our model is that our normalization model is able to better capture the structural information and contrast between the source and processed image, leading to normalized images of  higher quality. In turn, these enhanced normalized images enable better feature extraction and more accurate predictions.

Another possible factor contributing to the superior performance of our model in survival prediction tasks is the use of the Mean Teacher approach, which has been shown to be effective in improving the performance of deep neural networks in the case of limited annotated data. This approach involves training a student network to mimic the output of a more robust teacher network, resulting in improved accuracy and generalization.

We assessed our model by predicting the TRG values and comparing them to the actual gradings determined by a pathologist. Our model achieved an overall accuracy of 86\% in predicting TRG values, with a sensitivity of 80\% and a specificity of 91\%. This indicates that our model is capable of accurately predicting whether a patient's tumor will respond to neoadjuvant therapy, which is crucial for treatment planning.

The proposed model demonstrated superior performance compared to the comparative models in separating the low and high-risk patients based on the Kaplan-Meier curves. Specifically, our model achieved a higher c-index and a more distinct separation between the low and high-risk groups, indicating an improved trend in anticipating patient outcomes.

However, there are some limitations to this study that would need to be addressed in future work. The first limitation is the lack of diversity in the dataset used for evaluation, which may limit the generalizability of the model to other datasets. Another limitation is the omission of a comprehensive analysis of the hyperparameters of the proposed model, which could potentially impact the performance of our approach. Therefore, future studies could explore the generalizability of our approach with multi-centric datasets, as well as investigate the relevance of different hyperparameters which could provide further insight into the performance of the model.

The performance of ViT distilled models (KD-SSL, KD-TaT, KD-GasHis, and KD-PVT) was compared to other models in this study. Results showed that the KD-PVT was able to outperform the SSL model that served as a Teacher in the training process, achieving the highest performance in survival and time-to-recurrence prediction tasks, as demonstrated by Table \ref{tab7}. These results suggest that using ViT as a distilled model is a promising approach for synthesizing discriminative features in the context survival prediction in CLM histopathological image analysis and outperforms traditional models in this task.

Overall, the combination of the GAN normalization model with the survival prediction pipeline, integrating knowledge distillation demonstrated promising performance in both image feature extraction and survival prediction tasks, highlighting the potential of these approaches in further HPS analysis studies.

Our results demonstrate the potential of using DL to improve the prognostic prediction of CLM patients, which could greatly promote early detection in the management of liver cancer patients. The ability to automate the prognosis prediction process, not only increases the efficiency and reduces the labor-intensive and time-consuming aspects of traditional annotation methods, but also reduces the inter- and intra-observer variability. The proposed pipeline could help improve the prognostic prediction of CLM patients, which could ultimately lead to better treatment options and better outcomes for the patients.

\section{Conclusion}
To conclude, the proposed end-to-end approach for prognosis prediction, based on machine learning of WSI features and semi-supervised tissue classification, and knowledge distillation with Vision Transformer (ViT) achieved promising results in predicting patient prognosis for colorectal liver metastasis. Moreover, the model was able to predict TRG values with a high degree of accuracy, indicating its potential use in guiding treatment decisions. The proposed approach could provide automated prognosis information for pathologists and oncologists, and could greatly promote precision medicine progress in managing CLM patients. Future research will focus on evaluating the performance of the proposed pipeline on larger datasets and on its clinical implementation. Additionally, further analysis into the cellular level of the WSI could help extract features related to the cellular distribution and immunological infiltration that have been shown to be related to the prognosis.

\section{Acknowledgments}
Funding was provided by the Discovery program from the National Science and Engineering Research Council of Canada (NSERC). 
AT was supported by the IVADO PRF-1, Fonds de recherche du Québec-Santé (No. 34939).
ST was supported by the FRQ-S Young Clinician Scientist Seed Grant (No. 32633), the FRQS Clinician Scientist Junior-1\&2 Salary Award (No. 30861, No. 298832), the Institut du Cancer de Montr\'eal establishment award, and the Universit\'e de Montr\'eal Roger Des Groseillers Research Chair in Hepatopancreatobiliary Surgical Oncology. 
FA was supported by the TransMedTech Institute and its primary funding partner, the Canada First Research Excellence Fund.
We would like to thank L. Rousseau, S. Langevin and J. Bilodeau from the CHUM hepatopancreatobiliary biobank and prospective registry for patients recruitment, biospecimen acquisition, and maintenance of clinicopathological data.

\bibliographystyle{model2-names.bst}\biboptions{authoryear}
\bibliography{refs}

\begin{thebibliography}{64}
\expandafter\ifx\csname natexlab\endcsname\relax\def\natexlab#1{#1}\fi
\providecommand{\url}[1]{\texttt{#1}}
\providecommand{\href}[2]{#2}
\providecommand{\path}[1]{#1}
\providecommand{\DOIprefix}{doi:}
\providecommand{\ArXivprefix}{arXiv:}
\providecommand{\URLprefix}{URL: }
\providecommand{\Pubmedprefix}{pmid:}
\providecommand{\doi}[1]{\href{http://dx.doi.org/#1}{\path{#1}}}
\providecommand{\Pubmed}[1]{\href{pmid:#1}{\path{#1}}}
\providecommand{\bibinfo}[2]{#2}
\ifx\xfnm\relax \def\xfnm[#1]{\unskip,\space#1}\fi
\bibitem[{Alzubaidi et~al.(2021)Alzubaidi, Zhang, Humaidi, Al-Dujaili, Duan, Al-Shamma, Santamar{\'\i}a, Fadhel, Al-Amidie and Farhan}]{alzubaidi2021review}
\bibinfo{author}{Alzubaidi, L.}, \bibinfo{author}{Zhang, J.}, \bibinfo{author}{Humaidi, A.J.}, \bibinfo{author}{Al-Dujaili, A.}, \bibinfo{author}{Duan, Y.}, \bibinfo{author}{Al-Shamma, O.}, \bibinfo{author}{Santamar{\'\i}a, J.}, \bibinfo{author}{Fadhel, M.A.}, \bibinfo{author}{Al-Amidie, M.}, \bibinfo{author}{Farhan, L.}, \bibinfo{year}{2021}.
\newblock \bibinfo{title}{Review of deep learning: Concepts, cnn architectures, challenges, applications, future directions}.
\newblock \bibinfo{journal}{Journal of big Data} \bibinfo{volume}{8}, \bibinfo{pages}{1--74}.
\bibitem[{Bakurov et~al.(2022)Bakurov, Buzzelli, Schettini, Castelli and Vanneschi}]{bakurov2022structural}
\bibinfo{author}{Bakurov, I.}, \bibinfo{author}{Buzzelli, M.}, \bibinfo{author}{Schettini, R.}, \bibinfo{author}{Castelli, M.}, \bibinfo{author}{Vanneschi, L.}, \bibinfo{year}{2022}.
\newblock \bibinfo{title}{Structural similarity index (ssim) revisited: A data-driven approach}.
\newblock \bibinfo{journal}{Expert Systems with Applications} \bibinfo{volume}{189}, \bibinfo{pages}{116087}.
\bibitem[{Banegas-Luna et~al.(2021)Banegas-Luna, Pe{\~n}a-Garc{\'\i}a, Iftene, Guadagni, Ferroni, Scarpato, Zanzotto, Bueno-Crespo and P{\'e}rez-S{\'a}nchez}]{banegas2021towards}
\bibinfo{author}{Banegas-Luna, A.J.}, \bibinfo{author}{Pe{\~n}a-Garc{\'\i}a, J.}, \bibinfo{author}{Iftene, A.}, \bibinfo{author}{Guadagni, F.}, \bibinfo{author}{Ferroni, P.}, \bibinfo{author}{Scarpato, N.}, \bibinfo{author}{Zanzotto, F.M.}, \bibinfo{author}{Bueno-Crespo, A.}, \bibinfo{author}{P{\'e}rez-S{\'a}nchez, H.}, \bibinfo{year}{2021}.
\newblock \bibinfo{title}{Towards the interpretability of machine learning predictions for medical applications targeting personalised therapies: A cancer case survey}.
\newblock \bibinfo{journal}{International Journal of Molecular Sciences} \bibinfo{volume}{22}, \bibinfo{pages}{4394}.
\bibitem[{Boehm et~al.(2022)Boehm, Khosravi, Vanguri, Gao and Shah}]{boehm2022harnessing}
\bibinfo{author}{Boehm, K.M.}, \bibinfo{author}{Khosravi, P.}, \bibinfo{author}{Vanguri, R.}, \bibinfo{author}{Gao, J.}, \bibinfo{author}{Shah, S.P.}, \bibinfo{year}{2022}.
\newblock \bibinfo{title}{Harnessing multimodal data integration to advance precision oncology}.
\newblock \bibinfo{journal}{Nature Reviews Cancer} \bibinfo{volume}{22}, \bibinfo{pages}{114--126}.
\bibitem[{Border and Sarder(2022)}]{border2022growing}
\bibinfo{author}{Border, S.P.}, \bibinfo{author}{Sarder, P.}, \bibinfo{year}{2022}.
\newblock \bibinfo{title}{From what to why, the growing need for a focus shift toward explainability of ai in digital pathology}.
\newblock \bibinfo{journal}{Frontiers in Physiology} \bibinfo{volume}{12}, \bibinfo{pages}{2397}.
\bibitem[{Campanella et~al.(2019)Campanella, Hanna, Geneslaw, Miraflor, Werneck Krauss~Silva, Busam, Brogi, Reuter, Klimstra and Fuchs}]{campanella2019clinical}
\bibinfo{author}{Campanella, G.}, \bibinfo{author}{Hanna, M.G.}, \bibinfo{author}{Geneslaw, L.}, \bibinfo{author}{Miraflor, A.}, \bibinfo{author}{Werneck Krauss~Silva, V.}, \bibinfo{author}{Busam, K.J.}, \bibinfo{author}{Brogi, E.}, \bibinfo{author}{Reuter, V.E.}, \bibinfo{author}{Klimstra, D.S.}, \bibinfo{author}{Fuchs, T.J.}, \bibinfo{year}{2019}.
\newblock \bibinfo{title}{Clinical-grade computational pathology using weakly supervised deep learning on whole slide images}.
\newblock \bibinfo{journal}{Nature medicine} \bibinfo{volume}{25}, \bibinfo{pages}{1301--1309}.
\bibitem[{Chen et~al.(2022)Chen, Li, Wang, Li, Rahaman, Sun, Hu, Li, Liu, Sun et~al.}]{chen2022gashis}
\bibinfo{author}{Chen, H.}, \bibinfo{author}{Li, C.}, \bibinfo{author}{Wang, G.}, \bibinfo{author}{Li, X.}, \bibinfo{author}{Rahaman, M.M.}, \bibinfo{author}{Sun, H.}, \bibinfo{author}{Hu, W.}, \bibinfo{author}{Li, Y.}, \bibinfo{author}{Liu, W.}, \bibinfo{author}{Sun, C.}, et~al., \bibinfo{year}{2022}.
\newblock \bibinfo{title}{Gashis-transformer: A multi-scale visual transformer approach for gastric histopathological image detection}.
\newblock \bibinfo{journal}{Pattern Recognition} \bibinfo{volume}{130}, \bibinfo{pages}{108827}.
\bibitem[{Chen and Krishnan(2022)}]{chen2022self}
\bibinfo{author}{Chen, R.J.}, \bibinfo{author}{Krishnan, R.G.}, \bibinfo{year}{2022}.
\newblock \bibinfo{title}{Self-supervised vision transformers learn visual concepts in histopathology}.
\newblock \bibinfo{journal}{arXiv preprint arXiv:2203.00585} .
\bibitem[{Ciompi et~al.(2017)Ciompi, Geessink, Bejnordi, De~Souza, Baidoshvili, Litjens, Van~Ginneken, Nagtegaal and Van Der~Laak}]{ciompi2017importance}
\bibinfo{author}{Ciompi, F.}, \bibinfo{author}{Geessink, O.}, \bibinfo{author}{Bejnordi, B.E.}, \bibinfo{author}{De~Souza, G.S.}, \bibinfo{author}{Baidoshvili, A.}, \bibinfo{author}{Litjens, G.}, \bibinfo{author}{Van~Ginneken, B.}, \bibinfo{author}{Nagtegaal, I.}, \bibinfo{author}{Van Der~Laak, J.}, \bibinfo{year}{2017}.
\newblock \bibinfo{title}{The importance of stain normalization in colorectal tissue classification with convolutional networks}, in: \bibinfo{booktitle}{2017 IEEE 14th International Symposium on Biomedical Imaging (ISBI 2017)}, \bibinfo{organization}{IEEE}. pp. \bibinfo{pages}{160--163}.
\bibitem[{Cui and Zhang(2021)}]{cui2021artificial}
\bibinfo{author}{Cui, M.}, \bibinfo{author}{Zhang, D.Y.}, \bibinfo{year}{2021}.
\newblock \bibinfo{title}{Artificial intelligence and computational pathology}.
\newblock \bibinfo{journal}{Laboratory Investigation} \bibinfo{volume}{101}, \bibinfo{pages}{412--422}.
\bibitem[{Deepa and Gunavathi(2022)}]{deepa2022systematic}
\bibinfo{author}{Deepa, P.}, \bibinfo{author}{Gunavathi, C.}, \bibinfo{year}{2022}.
\newblock \bibinfo{title}{A systematic review on machine learning and deep learning techniques in cancer survival prediction}.
\newblock \bibinfo{journal}{Progress in Biophysics and Molecular Biology} .
\bibitem[{Deng et~al.(2009)Deng, Dong, Socher, Li, Li and Fei-Fei}]{deng2009imagenet}
\bibinfo{author}{Deng, J.}, \bibinfo{author}{Dong, W.}, \bibinfo{author}{Socher, R.}, \bibinfo{author}{Li, L.J.}, \bibinfo{author}{Li, K.}, \bibinfo{author}{Fei-Fei, L.}, \bibinfo{year}{2009}.
\newblock \bibinfo{title}{Imagenet: A large-scale hierarchical image database}, in: \bibinfo{booktitle}{2009 IEEE conference on computer vision and pattern recognition}, \bibinfo{organization}{Ieee}. pp. \bibinfo{pages}{248--255}.
\bibitem[{Dimitriou et~al.(2019)Dimitriou, Arandjelovi{\'c} and Caie}]{dimitriou2019deep}
\bibinfo{author}{Dimitriou, N.}, \bibinfo{author}{Arandjelovi{\'c}, O.}, \bibinfo{author}{Caie, P.D.}, \bibinfo{year}{2019}.
\newblock \bibinfo{title}{Deep learning for whole slide image analysis: an overview}.
\newblock \bibinfo{journal}{Frontiers in medicine} \bibinfo{volume}{6}, \bibinfo{pages}{264}.
\bibitem[{Dosovitskiy et~al.(2020)Dosovitskiy, Beyer, Kolesnikov, Weissenborn, Zhai, Unterthiner, Dehghani, Minderer, Heigold, Gelly et~al.}]{dosovitskiy2020image}
\bibinfo{author}{Dosovitskiy, A.}, \bibinfo{author}{Beyer, L.}, \bibinfo{author}{Kolesnikov, A.}, \bibinfo{author}{Weissenborn, D.}, \bibinfo{author}{Zhai, X.}, \bibinfo{author}{Unterthiner, T.}, \bibinfo{author}{Dehghani, M.}, \bibinfo{author}{Minderer, M.}, \bibinfo{author}{Heigold, G.}, \bibinfo{author}{Gelly, S.}, et~al., \bibinfo{year}{2020}.
\newblock \bibinfo{title}{An image is worth 16x16 words: Transformers for image recognition at scale}.
\newblock \bibinfo{journal}{arXiv preprint arXiv:2010.11929} .
\bibitem[{Elforaici et~al.(2022)Elforaici, Montagnon, Azzi, Trudel, Nguyen, Turcotte, Tang and Kadoury}]{elforaici2022semi}
\bibinfo{author}{Elforaici, M.E.A.}, \bibinfo{author}{Montagnon, E.}, \bibinfo{author}{Azzi, F.}, \bibinfo{author}{Trudel, D.}, \bibinfo{author}{Nguyen, B.}, \bibinfo{author}{Turcotte, S.}, \bibinfo{author}{Tang, A.}, \bibinfo{author}{Kadoury, S.}, \bibinfo{year}{2022}.
\newblock \bibinfo{title}{Semi-supervised tumor response grade classification from histology images of colorectal liver metastases}, in: \bibinfo{booktitle}{2022 IEEE 19th International Symposium on Biomedical Imaging (ISBI)}, \bibinfo{organization}{IEEE}. pp. \bibinfo{pages}{1--5}.
\bibitem[{Fong et~al.(1999)Fong, Fortner, Sun, Brennan and Blumgart}]{fong1999clinical}
\bibinfo{author}{Fong, Y.}, \bibinfo{author}{Fortner, J.}, \bibinfo{author}{Sun, R.L.}, \bibinfo{author}{Brennan, M.F.}, \bibinfo{author}{Blumgart, L.H.}, \bibinfo{year}{1999}.
\newblock \bibinfo{title}{Clinical score for predicting recurrence after hepatic resection for metastatic colorectal cancer: analysis of 1001 consecutive cases}.
\newblock \bibinfo{journal}{Annals of surgery} \bibinfo{volume}{230}, \bibinfo{pages}{309}.
\bibitem[{Goode et~al.(2013)Goode, Gilbert, Harkes, Jukic and Satyanarayanan}]{goode2013openslide}
\bibinfo{author}{Goode, A.}, \bibinfo{author}{Gilbert, B.}, \bibinfo{author}{Harkes, J.}, \bibinfo{author}{Jukic, D.}, \bibinfo{author}{Satyanarayanan, M.}, \bibinfo{year}{2013}.
\newblock \bibinfo{title}{Openslide: A vendor-neutral software foundation for digital pathology}.
\newblock \bibinfo{journal}{Journal of pathology informatics} \bibinfo{volume}{4}, \bibinfo{pages}{27}.
\bibitem[{Gou et~al.(2021)Gou, Yu, Maybank and Tao}]{gou2021knowledge}
\bibinfo{author}{Gou, J.}, \bibinfo{author}{Yu, B.}, \bibinfo{author}{Maybank, S.J.}, \bibinfo{author}{Tao, D.}, \bibinfo{year}{2021}.
\newblock \bibinfo{title}{Knowledge distillation: A survey}.
\newblock \bibinfo{journal}{International Journal of Computer Vision} \bibinfo{volume}{129}, \bibinfo{pages}{1789--1819}.
\bibitem[{Hady and Schwenker(2013)}]{hady2013semi}
\bibinfo{author}{Hady, M.F.A.}, \bibinfo{author}{Schwenker, F.}, \bibinfo{year}{2013}.
\newblock \bibinfo{title}{Semi-supervised learning}.
\newblock \bibinfo{journal}{Handbook on Neural Information Processing} , \bibinfo{pages}{215--239}.
\bibitem[{{Hamamatsu Photonics K.K.}()}]{Hamamatsu}
\bibinfo{author}{{Hamamatsu Photonics K.K.}}, .
\newblock \bibinfo{title}{Digital slide scanner}.
\newblock \bibinfo{howpublished}{\url{https://www.hamamatsu.com/jp/en/product/life-science-and-medical-systems/digital-slide-scanner.html}}.
\newblock \bibinfo{note}{Accessed: February 12, 2023}.
\bibitem[{{Hamamatsu Photonics K.K.}(2022)}]{NDPview2}
\bibinfo{author}{{Hamamatsu Photonics K.K.}}, \bibinfo{year}{2022}.
\newblock \bibinfo{title}{{NDP.view2}}.
\newblock \bibinfo{howpublished}{Software}.
\newblock \bibinfo{note}{February 12, 2023, from \url{https://www.hamamatsu.com/eu/en/product/type/U12388-01/index.html}}.
\bibitem[{Hatamizadeh et~al.(2022)Hatamizadeh, Yin, Kautz and Molchanov}]{hatamizadeh2022global}
\bibinfo{author}{Hatamizadeh, A.}, \bibinfo{author}{Yin, H.}, \bibinfo{author}{Kautz, J.}, \bibinfo{author}{Molchanov, P.}, \bibinfo{year}{2022}.
\newblock \bibinfo{title}{Global context vision transformers}.
\newblock \bibinfo{journal}{arXiv preprint arXiv:2206.09959} .
\bibitem[{Howard et~al.(2019)Howard, Sandler, Chu, Chen, Chen, Tan, Wang, Zhu, Pang, Vasudevan et~al.}]{howard2019searching}
\bibinfo{author}{Howard, A.}, \bibinfo{author}{Sandler, M.}, \bibinfo{author}{Chu, G.}, \bibinfo{author}{Chen, L.C.}, \bibinfo{author}{Chen, B.}, \bibinfo{author}{Tan, M.}, \bibinfo{author}{Wang, W.}, \bibinfo{author}{Zhu, Y.}, \bibinfo{author}{Pang, R.}, \bibinfo{author}{Vasudevan, V.}, et~al., \bibinfo{year}{2019}.
\newblock \bibinfo{title}{Searching for mobilenetv3}, in: \bibinfo{booktitle}{Proceedings of the IEEE/CVF international conference on computer vision}, pp. \bibinfo{pages}{1314--1324}.
\bibitem[{Ikromjanov et~al.(2022)Ikromjanov, Bhattacharjee, Hwang, Sumon, Kim and Choi}]{ikromjanov2022whole}
\bibinfo{author}{Ikromjanov, K.}, \bibinfo{author}{Bhattacharjee, S.}, \bibinfo{author}{Hwang, Y.B.}, \bibinfo{author}{Sumon, R.I.}, \bibinfo{author}{Kim, H.C.}, \bibinfo{author}{Choi, H.K.}, \bibinfo{year}{2022}.
\newblock \bibinfo{title}{Whole slide image analysis and detection of prostate cancer using vision transformers}, in: \bibinfo{booktitle}{2022 International Conference on Artificial Intelligence in Information and Communication (ICAIIC)}, \bibinfo{organization}{IEEE}. pp. \bibinfo{pages}{399--402}.
\bibitem[{Katzman et~al.(2018)Katzman, Shaham, Cloninger, Bates, Jiang and Kluger}]{katzman2018deepsurv}
\bibinfo{author}{Katzman, J.L.}, \bibinfo{author}{Shaham, U.}, \bibinfo{author}{Cloninger, A.}, \bibinfo{author}{Bates, J.}, \bibinfo{author}{Jiang, T.}, \bibinfo{author}{Kluger, Y.}, \bibinfo{year}{2018}.
\newblock \bibinfo{title}{Deepsurv: personalized treatment recommender system using a cox proportional hazards deep neural network}.
\newblock \bibinfo{journal}{BMC medical research methodology} \bibinfo{volume}{18}, \bibinfo{pages}{1--12}.
\bibitem[{Khened et~al.(2021)Khened, Kori, Rajkumar, Krishnamurthi and Srinivasan}]{khened2021generalized}
\bibinfo{author}{Khened, M.}, \bibinfo{author}{Kori, A.}, \bibinfo{author}{Rajkumar, H.}, \bibinfo{author}{Krishnamurthi, G.}, \bibinfo{author}{Srinivasan, B.}, \bibinfo{year}{2021}.
\newblock \bibinfo{title}{A generalized deep learning framework for whole-slide image segmentation and analysis}.
\newblock \bibinfo{journal}{Scientific reports} \bibinfo{volume}{11}, \bibinfo{pages}{1--14}.
\bibitem[{Van~der Laak et~al.(2021)Van~der Laak, Litjens and Ciompi}]{van2021deep}
\bibinfo{author}{Van~der Laak, J.}, \bibinfo{author}{Litjens, G.}, \bibinfo{author}{Ciompi, F.}, \bibinfo{year}{2021}.
\newblock \bibinfo{title}{Deep learning in histopathology: the path to the clinic}.
\newblock \bibinfo{journal}{Nature medicine} \bibinfo{volume}{27}, \bibinfo{pages}{775--784}.
\bibitem[{Li et~al.(2022)Li, Li, Rahaman, Sun, Li, Wu, Yao and Grzegorzek}]{li2022comprehensive}
\bibinfo{author}{Li, X.}, \bibinfo{author}{Li, C.}, \bibinfo{author}{Rahaman, M.M.}, \bibinfo{author}{Sun, H.}, \bibinfo{author}{Li, X.}, \bibinfo{author}{Wu, J.}, \bibinfo{author}{Yao, Y.}, \bibinfo{author}{Grzegorzek, M.}, \bibinfo{year}{2022}.
\newblock \bibinfo{title}{A comprehensive review of computer-aided whole-slide image analysis: from datasets to feature extraction, segmentation, classification and detection approaches}.
\newblock \bibinfo{journal}{Artificial Intelligence Review} \bibinfo{volume}{55}, \bibinfo{pages}{4809--4878}.
\bibitem[{Liang et~al.(2020)Liang, Plataniotis and Li}]{liang2020stain}
\bibinfo{author}{Liang, H.}, \bibinfo{author}{Plataniotis, K.N.}, \bibinfo{author}{Li, X.}, \bibinfo{year}{2020}.
\newblock \bibinfo{title}{Stain style transfer of histopathology images via structure-preserved generative learning}, in: \bibinfo{booktitle}{Machine Learning for Medical Image Reconstruction: Third International Workshop, MLMIR 2020, Held in Conjunction with MICCAI 2020, Lima, Peru, October 8, 2020, Proceedings 3}, \bibinfo{organization}{Springer}. pp. \bibinfo{pages}{153--162}.
\bibitem[{Lin et~al.(2022)Lin, Xie, Wang, Yu, Chang, Liang and Wang}]{lin2022knowledge}
\bibinfo{author}{Lin, S.}, \bibinfo{author}{Xie, H.}, \bibinfo{author}{Wang, B.}, \bibinfo{author}{Yu, K.}, \bibinfo{author}{Chang, X.}, \bibinfo{author}{Liang, X.}, \bibinfo{author}{Wang, G.}, \bibinfo{year}{2022}.
\newblock \bibinfo{title}{Knowledge distillation via the target-aware transformer}, in: \bibinfo{booktitle}{Proceedings of the IEEE/CVF Conference on Computer Vision and Pattern Recognition}, pp. \bibinfo{pages}{10915--10924}.
\bibitem[{Macenko et~al.(2009)Macenko, Niethammer, Marron, Borland, Woosley, Guan, Schmitt and Thomas}]{macenko2009method}
\bibinfo{author}{Macenko, M.}, \bibinfo{author}{Niethammer, M.}, \bibinfo{author}{Marron, J.S.}, \bibinfo{author}{Borland, D.}, \bibinfo{author}{Woosley, J.T.}, \bibinfo{author}{Guan, X.}, \bibinfo{author}{Schmitt, C.}, \bibinfo{author}{Thomas, N.E.}, \bibinfo{year}{2009}.
\newblock \bibinfo{title}{A method for normalizing histology slides for quantitative analysis}, in: \bibinfo{booktitle}{2009 IEEE international symposium on biomedical imaging: from nano to macro}, \bibinfo{organization}{IEEE}. pp. \bibinfo{pages}{1107--1110}.
\bibitem[{Mandard et~al.(1994)Mandard, Dalibard, Mandard, Marnay, Henry-Amar, Petiot, Roussel, Jacob, Segol, Samama et~al.}]{mandard1994pathologic}
\bibinfo{author}{Mandard, A.M.}, \bibinfo{author}{Dalibard, F.}, \bibinfo{author}{Mandard, J.C.}, \bibinfo{author}{Marnay, J.}, \bibinfo{author}{Henry-Amar, M.}, \bibinfo{author}{Petiot, J.F.}, \bibinfo{author}{Roussel, A.}, \bibinfo{author}{Jacob, J.H.}, \bibinfo{author}{Segol, P.}, \bibinfo{author}{Samama, G.}, et~al., \bibinfo{year}{1994}.
\newblock \bibinfo{title}{Pathologic assessment of tumor regression after preoperative chemoradiotherapy of esophageal carcinoma. clinicopathologic correlations}.
\newblock \bibinfo{journal}{Cancer} \bibinfo{volume}{73}, \bibinfo{pages}{2680--2686}.
\bibitem[{McQuin et~al.(2018)McQuin, Goodman, Chernyshev, Kamentsky, Cimini, Karhohs, Doan, Ding, Rafelski, Thirstrup et~al.}]{mcquin2018cellprofiler}
\bibinfo{author}{McQuin, C.}, \bibinfo{author}{Goodman, A.}, \bibinfo{author}{Chernyshev, V.}, \bibinfo{author}{Kamentsky, L.}, \bibinfo{author}{Cimini, B.A.}, \bibinfo{author}{Karhohs, K.W.}, \bibinfo{author}{Doan, M.}, \bibinfo{author}{Ding, L.}, \bibinfo{author}{Rafelski, S.M.}, \bibinfo{author}{Thirstrup, D.}, et~al., \bibinfo{year}{2018}.
\newblock \bibinfo{title}{Cellprofiler 3.0: Next-generation image processing for biology}.
\newblock \bibinfo{journal}{PLoS biology} \bibinfo{volume}{16}, \bibinfo{pages}{e2005970}.
\bibitem[{Miyato et~al.(2018)Miyato, Maeda, Koyama and Ishii}]{miyato2018virtual}
\bibinfo{author}{Miyato, T.}, \bibinfo{author}{Maeda, S.i.}, \bibinfo{author}{Koyama, M.}, \bibinfo{author}{Ishii, S.}, \bibinfo{year}{2018}.
\newblock \bibinfo{title}{Virtual adversarial training: a regularization method for supervised and semi-supervised learning}.
\newblock \bibinfo{journal}{IEEE transactions on pattern analysis and machine intelligence} \bibinfo{volume}{41}, \bibinfo{pages}{1979--1993}.
\bibitem[{Reinhard et~al.(2001)Reinhard, Adhikhmin, Gooch and Shirley}]{reinhard2001color}
\bibinfo{author}{Reinhard, E.}, \bibinfo{author}{Adhikhmin, M.}, \bibinfo{author}{Gooch, B.}, \bibinfo{author}{Shirley, P.}, \bibinfo{year}{2001}.
\newblock \bibinfo{title}{Color transfer between images}.
\newblock \bibinfo{journal}{IEEE Computer graphics and applications} \bibinfo{volume}{21}, \bibinfo{pages}{34--41}.
\bibitem[{Ren et~al.(2023)Ren, Jia, Zhao, Qiang, Wu, Han, Zhao and Sun}]{ren2023weakly}
\bibinfo{author}{Ren, X.}, \bibinfo{author}{Jia, L.}, \bibinfo{author}{Zhao, Z.}, \bibinfo{author}{Qiang, Y.}, \bibinfo{author}{Wu, W.}, \bibinfo{author}{Han, P.}, \bibinfo{author}{Zhao, J.}, \bibinfo{author}{Sun, J.}, \bibinfo{year}{2023}.
\newblock \bibinfo{title}{Weakly supervised label propagation algorithm classifies lung cancer imaging subtypes}.
\newblock \bibinfo{journal}{Scientific Reports} \bibinfo{volume}{13}, \bibinfo{pages}{5167}.
\bibitem[{Rodgers and Nicewander(1988)}]{rodgers1988thirteen}
\bibinfo{author}{Rodgers, J.L.}, \bibinfo{author}{Nicewander, W.A.}, \bibinfo{year}{1988}.
\newblock \bibinfo{title}{Thirteen ways to look at the correlation coefficient}.
\newblock \bibinfo{journal}{American statistician} , \bibinfo{pages}{59--66}.
\bibitem[{Romero et~al.(2014)Romero, Ballas, Kahou, Chassang, Gatta and Bengio}]{romero2014fitnets}
\bibinfo{author}{Romero, A.}, \bibinfo{author}{Ballas, N.}, \bibinfo{author}{Kahou, S.E.}, \bibinfo{author}{Chassang, A.}, \bibinfo{author}{Gatta, C.}, \bibinfo{author}{Bengio, Y.}, \bibinfo{year}{2014}.
\newblock \bibinfo{title}{Fitnets: Hints for thin deep nets}.
\newblock \bibinfo{journal}{arXiv preprint arXiv:1412.6550} .
\bibitem[{Roth et~al.(2021)Roth, Yang, Xu, Wang and Xu}]{roth2021going}
\bibinfo{author}{Roth, H.R.}, \bibinfo{author}{Yang, D.}, \bibinfo{author}{Xu, Z.}, \bibinfo{author}{Wang, X.}, \bibinfo{author}{Xu, D.}, \bibinfo{year}{2021}.
\newblock \bibinfo{title}{Going to extremes: weakly supervised medical image segmentation}.
\newblock \bibinfo{journal}{Machine Learning and Knowledge Extraction} \bibinfo{volume}{3}, \bibinfo{pages}{507--524}.
\bibitem[{Rubbia-Brandt et~al.(2007)Rubbia-Brandt, Giostra, Brezault, Roth, Andres, Audard, Sartoretti, Dousset, Majno, Soubrane et~al.}]{rubbia2007importance}
\bibinfo{author}{Rubbia-Brandt, L.}, \bibinfo{author}{Giostra, E.}, \bibinfo{author}{Brezault, C.}, \bibinfo{author}{Roth, A.}, \bibinfo{author}{Andres, A.}, \bibinfo{author}{Audard, V.}, \bibinfo{author}{Sartoretti, P.}, \bibinfo{author}{Dousset, B.}, \bibinfo{author}{Majno, P.}, \bibinfo{author}{Soubrane, O.}, et~al., \bibinfo{year}{2007}.
\newblock \bibinfo{title}{Importance of histological tumor response assessment in predicting the outcome in patients with colorectal liver metastases treated with neo-adjuvant chemotherapy followed by liver surgery}.
\newblock \bibinfo{journal}{Annals of oncology} \bibinfo{volume}{18}, \bibinfo{pages}{299--304}.
\bibitem[{Runz et~al.(2021)Runz, Rusche, Schmidt, Weihrauch, Hesser and Weis}]{runz2021normalization}
\bibinfo{author}{Runz, M.}, \bibinfo{author}{Rusche, D.}, \bibinfo{author}{Schmidt, S.}, \bibinfo{author}{Weihrauch, M.R.}, \bibinfo{author}{Hesser, J.}, \bibinfo{author}{Weis, C.A.}, \bibinfo{year}{2021}.
\newblock \bibinfo{title}{Normalization of he-stained histological images using cycle consistent generative adversarial networks}.
\newblock \bibinfo{journal}{Diagnostic Pathology} \bibinfo{volume}{16}, \bibinfo{pages}{1--10}.
\bibitem[{Salvi et~al.(2020)Salvi, Michielli and Molinari}]{salvi2020stain}
\bibinfo{author}{Salvi, M.}, \bibinfo{author}{Michielli, N.}, \bibinfo{author}{Molinari, F.}, \bibinfo{year}{2020}.
\newblock \bibinfo{title}{Stain color adaptive normalization (scan) algorithm: Separation and standardization of histological stains in digital pathology}.
\newblock \bibinfo{journal}{Computer methods and programs in biomedicine} \bibinfo{volume}{193}, \bibinfo{pages}{105506}.
\bibitem[{Sandler et~al.(2018)Sandler, Howard, Zhu, Zhmoginov and Chen}]{sandler2018mobilenetv2}
\bibinfo{author}{Sandler, M.}, \bibinfo{author}{Howard, A.}, \bibinfo{author}{Zhu, M.}, \bibinfo{author}{Zhmoginov, A.}, \bibinfo{author}{Chen, L.C.}, \bibinfo{year}{2018}.
\newblock \bibinfo{title}{Mobilenetv2: Inverted residuals and linear bottlenecks}, in: \bibinfo{booktitle}{Proceedings of the IEEE conference on computer vision and pattern recognition}, pp. \bibinfo{pages}{4510--4520}.
\bibitem[{Shaw et~al.(2020)Shaw, Pajak, Lisowska, Tsaftaris and O'Neil}]{shaw2020teacher}
\bibinfo{author}{Shaw, S.}, \bibinfo{author}{Pajak, M.}, \bibinfo{author}{Lisowska, A.}, \bibinfo{author}{Tsaftaris, S.A.}, \bibinfo{author}{O'Neil, A.Q.}, \bibinfo{year}{2020}.
\newblock \bibinfo{title}{Teacher-student chain for efficient semi-supervised histology image classification}.
\newblock \bibinfo{journal}{arXiv preprint arXiv:2003.08797} .
\bibitem[{Tarvainen and Valpola(2017)}]{tarvainen2017mean}
\bibinfo{author}{Tarvainen, A.}, \bibinfo{author}{Valpola, H.}, \bibinfo{year}{2017}.
\newblock \bibinfo{title}{Mean teachers are better role models: Weight-averaged consistency targets improve semi-supervised deep learning results}.
\newblock \bibinfo{journal}{Advances in neural information processing systems} \bibinfo{volume}{30}.
\bibitem[{Tian et~al.(2019)Tian, Krishnan and Isola}]{tian2019contrastive}
\bibinfo{author}{Tian, Y.}, \bibinfo{author}{Krishnan, D.}, \bibinfo{author}{Isola, P.}, \bibinfo{year}{2019}.
\newblock \bibinfo{title}{Contrastive representation distillation}.
\newblock \bibinfo{journal}{arXiv preprint arXiv:1910.10699} .
\bibitem[{Vale-Silva and Rohr(2021)}]{vale2021long}
\bibinfo{author}{Vale-Silva, L.A.}, \bibinfo{author}{Rohr, K.}, \bibinfo{year}{2021}.
\newblock \bibinfo{title}{Long-term cancer survival prediction using multimodal deep learning}.
\newblock \bibinfo{journal}{Scientific Reports} \bibinfo{volume}{11}, \bibinfo{pages}{13505}.
\bibitem[{Vecchio et~al.(2005)Vecchio, Valentini, Minsky, Padula, Venkatraman, Balducci, Miccich{\`e}, Ricci, Morganti, Gambacorta et~al.}]{vecchio2005relationship}
\bibinfo{author}{Vecchio, F.M.}, \bibinfo{author}{Valentini, V.}, \bibinfo{author}{Minsky, B.D.}, \bibinfo{author}{Padula, G.D.}, \bibinfo{author}{Venkatraman, E.S.}, \bibinfo{author}{Balducci, M.}, \bibinfo{author}{Miccich{\`e}, F.}, \bibinfo{author}{Ricci, R.}, \bibinfo{author}{Morganti, A.G.}, \bibinfo{author}{Gambacorta, M.A.}, et~al., \bibinfo{year}{2005}.
\newblock \bibinfo{title}{The relationship of pathologic tumor regression grade (trg) and outcomes after preoperative therapy in rectal cancer}.
\newblock \bibinfo{journal}{International Journal of Radiation Oncology* Biology* Physics} \bibinfo{volume}{62}, \bibinfo{pages}{752--760}.
\bibitem[{Vijh et~al.(2021)Vijh, Saraswat and Kumar}]{vijh2021new}
\bibinfo{author}{Vijh, S.}, \bibinfo{author}{Saraswat, M.}, \bibinfo{author}{Kumar, S.}, \bibinfo{year}{2021}.
\newblock \bibinfo{title}{A new complete color normalization method for h\&e stained histopatholgical images}.
\newblock \bibinfo{journal}{Applied Intelligence} , \bibinfo{pages}{1--14}.
\bibitem[{Wang et~al.(2021)Wang, Xie, Li, Fan, Song, Liang, Lu, Luo and Shao}]{wang2021pyramid}
\bibinfo{author}{Wang, W.}, \bibinfo{author}{Xie, E.}, \bibinfo{author}{Li, X.}, \bibinfo{author}{Fan, D.P.}, \bibinfo{author}{Song, K.}, \bibinfo{author}{Liang, D.}, \bibinfo{author}{Lu, T.}, \bibinfo{author}{Luo, P.}, \bibinfo{author}{Shao, L.}, \bibinfo{year}{2021}.
\newblock \bibinfo{title}{Pyramid vision transformer: A versatile backbone for dense prediction without convolutions}, in: \bibinfo{booktitle}{Proceedings of the IEEE/CVF international conference on computer vision}, pp. \bibinfo{pages}{568--578}.
\bibitem[{Wang et~al.(2019)Wang, Chen, Gan, Lin, Dou, Tsougenis, Huang, Cai and Heng}]{wang2019weakly}
\bibinfo{author}{Wang, X.}, \bibinfo{author}{Chen, H.}, \bibinfo{author}{Gan, C.}, \bibinfo{author}{Lin, H.}, \bibinfo{author}{Dou, Q.}, \bibinfo{author}{Tsougenis, E.}, \bibinfo{author}{Huang, Q.}, \bibinfo{author}{Cai, M.}, \bibinfo{author}{Heng, P.A.}, \bibinfo{year}{2019}.
\newblock \bibinfo{title}{Weakly supervised deep learning for whole slide lung cancer image analysis}.
\newblock \bibinfo{journal}{IEEE transactions on cybernetics} \bibinfo{volume}{50}, \bibinfo{pages}{3950--3962}.
\bibitem[{Wang et~al.(2004)Wang, Bovik, Sheikh and Simoncelli}]{wang2004image}
\bibinfo{author}{Wang, Z.}, \bibinfo{author}{Bovik, A.C.}, \bibinfo{author}{Sheikh, H.R.}, \bibinfo{author}{Simoncelli, E.P.}, \bibinfo{year}{2004}.
\newblock \bibinfo{title}{Image quality assessment: from error visibility to structural similarity}.
\newblock \bibinfo{journal}{IEEE transactions on image processing} \bibinfo{volume}{13}, \bibinfo{pages}{600--612}.
\bibitem[{Wulczyn et~al.(2021)Wulczyn, Steiner, Moran, Plass, Reihs, Tan, Flament-Auvigne, Brown, Regitnig, Chen et~al.}]{wulczyn2021interpretable}
\bibinfo{author}{Wulczyn, E.}, \bibinfo{author}{Steiner, D.F.}, \bibinfo{author}{Moran, M.}, \bibinfo{author}{Plass, M.}, \bibinfo{author}{Reihs, R.}, \bibinfo{author}{Tan, F.}, \bibinfo{author}{Flament-Auvigne, I.}, \bibinfo{author}{Brown, T.}, \bibinfo{author}{Regitnig, P.}, \bibinfo{author}{Chen, P.H.C.}, et~al., \bibinfo{year}{2021}.
\newblock \bibinfo{title}{Interpretable survival prediction for colorectal cancer using deep learning}.
\newblock \bibinfo{journal}{NPJ digital medicine} \bibinfo{volume}{4}, \bibinfo{pages}{71}.
\bibitem[{Wulczyn et~al.(2020)Wulczyn, Steiner, Xu, Sadhwani, Wang, Flament-Auvigne, Mermel, Chen, Liu and Stumpe}]{wulczyn2020deep}
\bibinfo{author}{Wulczyn, E.}, \bibinfo{author}{Steiner, D.F.}, \bibinfo{author}{Xu, Z.}, \bibinfo{author}{Sadhwani, A.}, \bibinfo{author}{Wang, H.}, \bibinfo{author}{Flament-Auvigne, I.}, \bibinfo{author}{Mermel, C.H.}, \bibinfo{author}{Chen, P.H.C.}, \bibinfo{author}{Liu, Y.}, \bibinfo{author}{Stumpe, M.C.}, \bibinfo{year}{2020}.
\newblock \bibinfo{title}{Deep learning-based survival prediction for multiple cancer types using histopathology images}.
\newblock \bibinfo{journal}{PloS one} \bibinfo{volume}{15}, \bibinfo{pages}{e0233678}.
\bibitem[{Xi and Xu(2021)}]{xi2021global}
\bibinfo{author}{Xi, Y.}, \bibinfo{author}{Xu, P.}, \bibinfo{year}{2021}.
\newblock \bibinfo{title}{Global colorectal cancer burden in 2020 and projections to 2040}.
\newblock \bibinfo{journal}{Translational oncology} \bibinfo{volume}{14}, \bibinfo{pages}{101174}.
\bibitem[{Yalniz et~al.(2019)Yalniz, J{\'e}gou, Chen, Paluri and Mahajan}]{yalniz2019billion}
\bibinfo{author}{Yalniz, I.Z.}, \bibinfo{author}{J{\'e}gou, H.}, \bibinfo{author}{Chen, K.}, \bibinfo{author}{Paluri, M.}, \bibinfo{author}{Mahajan, D.}, \bibinfo{year}{2019}.
\newblock \bibinfo{title}{Billion-scale semi-supervised learning for image classification}.
\newblock \bibinfo{journal}{arXiv preprint arXiv:1905.00546} .
\bibitem[{Yang et~al.(2017)Yang, Tianyi~Zhou, Cai and Soon~Ong}]{yang2017miml}
\bibinfo{author}{Yang, H.}, \bibinfo{author}{Tianyi~Zhou, J.}, \bibinfo{author}{Cai, J.}, \bibinfo{author}{Soon~Ong, Y.}, \bibinfo{year}{2017}.
\newblock \bibinfo{title}{Miml-fcn+: Multi-instance multi-label learning via fully convolutional networks with privileged information}, in: \bibinfo{booktitle}{Proceedings of the IEEE conference on computer vision and pattern recognition}, pp. \bibinfo{pages}{1577--1585}.
\bibitem[{Yang et~al.(2022)Yang, Li, Zeng, Li, Yuan and Li}]{yang2022vitkd}
\bibinfo{author}{Yang, Z.}, \bibinfo{author}{Li, Z.}, \bibinfo{author}{Zeng, A.}, \bibinfo{author}{Li, Z.}, \bibinfo{author}{Yuan, C.}, \bibinfo{author}{Li, Y.}, \bibinfo{year}{2022}.
\newblock \bibinfo{title}{Vitkd: Practical guidelines for vit feature knowledge distillation}.
\newblock \bibinfo{journal}{arXiv preprint arXiv:2209.02432} .
\bibitem[{Yao et~al.(2020)Yao, Zhu, Jonnagaddala, Hawkins and Huang}]{yao2020whole}
\bibinfo{author}{Yao, J.}, \bibinfo{author}{Zhu, X.}, \bibinfo{author}{Jonnagaddala, J.}, \bibinfo{author}{Hawkins, N.}, \bibinfo{author}{Huang, J.}, \bibinfo{year}{2020}.
\newblock \bibinfo{title}{Whole slide images based cancer survival prediction using attention guided deep multiple instance learning networks}.
\newblock \bibinfo{journal}{Medical Image Analysis} \bibinfo{volume}{65}, \bibinfo{pages}{101789}.
\bibitem[{Yu et~al.(2016)Yu, Zhang, Berry, Altman, R{\'e}, Rubin and Snyder}]{yu2016predicting}
\bibinfo{author}{Yu, K.H.}, \bibinfo{author}{Zhang, C.}, \bibinfo{author}{Berry, G.J.}, \bibinfo{author}{Altman, R.B.}, \bibinfo{author}{R{\'e}, C.}, \bibinfo{author}{Rubin, D.L.}, \bibinfo{author}{Snyder, M.}, \bibinfo{year}{2016}.
\newblock \bibinfo{title}{Predicting non-small cell lung cancer prognosis by fully automated microscopic pathology image features}.
\newblock \bibinfo{journal}{Nature communications} \bibinfo{volume}{7}, \bibinfo{pages}{12474}.
\bibitem[{Zheng et~al.(2022)Zheng, Wang and Yuan}]{zheng2022boosting}
\bibinfo{author}{Zheng, K.}, \bibinfo{author}{Wang, Y.}, \bibinfo{author}{Yuan, Y.}, \bibinfo{year}{2022}.
\newblock \bibinfo{title}{Boosting contrastive learning with relation knowledge distillation}, in: \bibinfo{booktitle}{Proceedings of the AAAI Conference on Artificial Intelligence}, pp. \bibinfo{pages}{3508--3516}.
\bibitem[{Zheng et~al.(2020)Zheng, Jiang, Zhang, Xie, Hu, Sun, Shi and Xue}]{zheng2020stain}
\bibinfo{author}{Zheng, Y.}, \bibinfo{author}{Jiang, Z.}, \bibinfo{author}{Zhang, H.}, \bibinfo{author}{Xie, F.}, \bibinfo{author}{Hu, D.}, \bibinfo{author}{Sun, S.}, \bibinfo{author}{Shi, J.}, \bibinfo{author}{Xue, C.}, \bibinfo{year}{2020}.
\newblock \bibinfo{title}{Stain standardization capsule for application-driven histopathological image normalization}.
\newblock \bibinfo{journal}{IEEE journal of biomedical and health informatics} \bibinfo{volume}{25}, \bibinfo{pages}{337--347}.
\bibitem[{Zhu et~al.(2016)Zhu, Yao and Huang}]{zhu2016deep}
\bibinfo{author}{Zhu, X.}, \bibinfo{author}{Yao, J.}, \bibinfo{author}{Huang, J.}, \bibinfo{year}{2016}.
\newblock \bibinfo{title}{Deep convolutional neural network for survival analysis with pathological images}, in: \bibinfo{booktitle}{2016 IEEE International Conference on Bioinformatics and Biomedicine (BIBM)}, \bibinfo{organization}{IEEE}. pp. \bibinfo{pages}{544--547}.
\bibitem[{Zhu et~al.(2017)Zhu, Yao, Zhu and Huang}]{zhu2017wsisa}
\bibinfo{author}{Zhu, X.}, \bibinfo{author}{Yao, J.}, \bibinfo{author}{Zhu, F.}, \bibinfo{author}{Huang, J.}, \bibinfo{year}{2017}.
\newblock \bibinfo{title}{Wsisa: Making survival prediction from whole slide histopathological images}, in: \bibinfo{booktitle}{Proceedings of the IEEE conference on computer vision and pattern recognition}, pp. \bibinfo{pages}{7234--7242}.

\end{thebibliography}

\end{document}